%% file: paper_v9.tex
\documentclass[sigconf]{acmart}

\hypersetup{draft}

\usepackage{flushend} % In  the  last  page  of  your  paper,  the  last  two  columns  must  be  of  equal  size

\usepackage{bm}
\usepackage{booktabs}
\usepackage[font=scriptsize]{subcaption}
\usepackage[font=footnotesize]{caption}

\usepackage[nowarn,acronyms,nonumberlist,nopostdot,nomain,nogroupskip]{glossaries}
\glsdisablehyper
\input{acronyms.tex}

\usepackage[capitalize]{cleveref}
\crefname{section}{Section}{Sections}
\crefname{figure}{Figure}{Figures}

\usepackage{tikz}
\usepackage{pgfplots}
\usepackage{tikzscale}
\usepackage{tikz-qtree}
\input{myTikz.tex}

\DeclareFontFamily{\encodingdefault}{\ttdefault}{\hyphenchar\font=`\-}

\AtBeginDocument{%
  \providecommand\BibTeX{{%
    \normalfont B\kern-0.5em{\scshape i\kern-0.25em b}\kern-0.8em\TeX}}}

\copyrightyear{2021}
\acmYear{2021}
\setcopyright{acmcopyright}\acmConference[WNS3 2021]{2021 Workshop on ns-3}{June 23--24, 2021}{Virtual Event, USA}
\acmBooktitle{2021 Workshop on ns-3 (WNS3 2021), June 23--24, 2021, Virtual Event, USA}
\acmPrice{15.00}
\acmDOI{10.1145/3460797.3460807}
\acmISBN{978-1-4503-9034-7/21/06}

\begin{document}
\pagestyle{empty} % Remove  the  headers  in  the  paper.  When  the  title  of  the  paper  is  long,  the  title  and  the  information of the conference will overlap in the headers. To avoid this problem, please remove the headers in the paper. 

\title{An ns-3 Implementation of a Bursty Traffic Framework\\for Virtual Reality Sources}

\author{Mattia Lecci}
\orcid{0000-0001-6347-1932}
\affiliation{%
  \institution{Dept. of Information Engineering\\University of Padova}
  \city{Padova}
  \country{Italy}
}
\email{leccimat@dei.unipd.it}

\author{Andrea Zanella}
\affiliation{%
  \institution{Dept. of Information Engineering\\University of Padova}
  \city{Padova}
  \country{Italy}
}
\email{zanella@dei.unipd.it}

\author{Michele Zorzi}
\affiliation{%
  \institution{Dept. of Information Engineering\\University of Padova}
  \city{Padova}
  \country{Italy}
}
\email{zorzi@dei.unipd.it}


\begin{abstract}
  Next-generation wireless communication technologies will allow users to obtain unprecedented performance, paving the way to new and immersive applications.
  A prominent application requiring high data rates and low communication delay is \gls{vr}, whose presence will become increasingly stronger in the years to come.
  To the best of our knowledge, we propose the first traffic model for \gls{vr} applications based on traffic traces acquired from a commercial \gls{vr} streaming software, allowing the community to further study and improve the technology to manage this type of traffic.
  This work implements ns-3 applications able to generate and process large bursts of packets, enabling the possibility of analyzing APP-level end-to-end metrics, making the source code, as well as the acquired \gls{vr} traffic traces, publicly available and open-source.
\end{abstract}

%%\begin{CCSXML}



\maketitle

\glsresetall
\glsunset{wifi}

\section{Introduction} % (fold)
\label{sec:introduction}

The growing demand for high-performance telecommunication networks is driving both industry and academia to push the boundaries of the achievable performance.

The \gls{itu} proposes requirements for \gls{imt2020} \gls{embb} such as peak \gls{dl} data rate of 20~Gbps with 4~ms user plane latency~\cite{ituMinimumRequirements}, for example by exploiting the large bandwidth available in the \gls{mmw} spectrum.
Similarly, \glspl{wlan} are also harvesting the potential of the \gls{mmw} band with a family of standards known as \gls{wigig}, including IEEE 802.11ad and 802.11ay.
While the former, first standardized in 2012~\cite{standard802.11ad} and later revised in 2016~\cite{standard802.11_2016}, is able to reach bit rates up to 8~Gbps, the latter is close to being officially standardized~\cite{tgayWebsite} and promises bit rates up to 100~Gbps.

These specifications for wireless systems enable a new generation of demanding applications such as high-definition wireless monitor, \gls{xr} headsets and other high-end wearables, data center inter-rack connectivity, wireless backhauling, and office docking, among others~\cite{tgayUsageModel}.

In particular, \gls{xr}, an umbrella name including technologies such as \gls{vr} and \gls{ar}, has been targeted as a key application with growing interest in the consumer market~\cite{huaweiVrArWhitePaper}.
Compact and portable devices with limited battery and computing power should be enabled to wirelessly support this type of demanding applications, to provide a fully immersive and realistic user experience.

To reach the limits of human vision, monitors with a resolution of 5073$\times$5707 per eye with 120~FPS refresh rate will be needed~\cite{huaweiVrArWhitePaper}.
These specifications suggest that rendering might be offloaded to a separate server as head-mounted displays should be light, silent, and comfortable enough to be worn for long periods of time.
This poses a significant strain on the wireless connection, requiring $\sim$167~Gbps of uncompressed video stream.
Clearly, real-time 360 video compression techniques allow to largely reduce these throughput requirements down to the order of 100--1000~Mbps, at the cost of some processing delay.

While throughput requirements are already very demanding, low latencies are the key to the success or the failure of \gls{xr} applications.
In fact, many studies showed that users tend to experience what is called \textit{motion} or \textit{cyber sickness} when their actions do not correspond to rapid reactions in the virtual world, causing disorientation and dizziness~\cite{huaweiVrArWhitePaper,hettinger1992motionSickness,groen2008motionSickness,vonMammen2016cyberSick,kim2017vrSickness}.
\textit{Motion-to-photon} latency is thus required to be at most 20~ms, translating into a network latency for video frames of 5--9~ms~\cite{tgayUsageModel,huaweiVrArWhitePaper}.

In its simplest and most ideal form, raw \gls{xr} traffic with a fixed frame rate $F$ could be modeled by periodic traffic, with period $1/F$ and constant frame size $S$ proportional to the display resolution.
Real traffic, though, is first of all compressed with one of the many existing video codecs, and then properly optimized for real-time low-latency streaming, resulting in encoded video frames of variable size.
Furthermore, the complexity of a given scene can also affect the time required to render it as well as the obtainable compression factor.
Together, these factors make both the video frame size and the period random, possibly correlated both in time and in the frame-size/period domains.

Given the interest of both industry and consumers in \gls{xr} applications and the peculiarity of the generated traffic, networks and networking protocols might be optimized to better support this type of traffic, ensuring its strict \gls{qos} requirements.

To the best of our knowledge, no prior work on \gls{xr} traffic modeling exists.
Cloud gaming~\cite{cai2016surveyCloudGaming} was identified as a closely related problem, where a remote server renders and streams a video to a client with limited computational resources, which only feeds basic information to the rendering servers, such as keys pressed and mouse movements.
The main difference with the problem under analysis is given by the more restrictive \gls{qos} constraints of \gls{xr} applications, mainly due to the limits imposed by motion sickness.
Furthermore, in cloud gaming, client and server are often in different \glspl{wlan}, making it harder to obtain reliable measurements of packet generation times.
In fact, due to the specific constraints and requirements of \gls{xr} applications, we expect the rendering server to be in a local network rather than being remotely accessed via the Internet.

Most works in the literature focus on network performance and limitations of cloud gaming~\cite{xue2015playingMeasurement}, and we could find only two main contributions addressing traffic analysis and modeling.
The authors of~\cite{claypool2014onlivePerformance} provide a simple traffic analysis for three different games played on \textit{OnLive}, a cloud gaming application that was shut down in 2015.
The analysis focuses on packet-level statistics, such as packet size, inter-packet time, and bit rate.
They measured the performance of the streaming service under speed-limited networks, showing an evident frame rate variability.
In~\cite{manzano2014onliveTraffic}, the authors tried to model the traffic generated by two games, also played on the \textit{OnLive} platform.
In particular, they recognized that video frames were split into multiple fragments, and re-aggregated them before studying their statistics.
A number of \gls{dl} and \gls{ul} data flows were recognized, and characterized in terms of \textit{application packet data unit size} and \textit{packet inter arrival time}.
Unfortunately, correlation among successive video frames was not modeled and the analysis referred to a single game played with an average data rate of about 5~Mbps.

The novelty of this paper can be summarized as follows:
\begin{enumerate}
  \item To the best of our knowledge, this is the first generative model for APP-level \gls{xr} traffic based on over 90~minutes of acquired and processed \gls{vr} traffic, with adaptable data rate and frame rate;
  \item we provide a flexible \gls{ns3} module for simulating applications with bursty behavior able to characterize both fragment-level and burst-level performance;
  \item we provide an implementation of the proposed \gls{xr} traffic model, as well as a trace-based model, on the bursty application framework;
  \item as a side contribution, several acquired \gls{vr} traffic traces are made available, allowing researchers to (i) use real \gls{vr} traffic in their simulations and (ii) further analyze and improve \gls{vr} traffic models.
\end{enumerate}

In the remainder of this work, we will describe the traffic acquisition and analysis in \cref{sec:vr_traffic_acquisition_and_analysis}, propose a flexible \gls{xr} traffic model in \cref{sec:traffic_model}, discuss about the \gls{ns3} implementation of the \textit{bursty application} framework in \cref{sec:ns_3_implementation}, validate the model and show a possible use case in \cref{sec:model_validation_and_possible_use_cases}, and finally draw the conclusions of this work and propose future works in \cref{sec:conclusions}.

% section introduction (end)

\section{VR Traffic: Acquisition and Analysis} % (fold)
\label{sec:vr_traffic_acquisition_and_analysis}

The traffic model that will be described in \cref{sec:traffic_model} is based on a set of acquisitions of \gls{vr} traffic.
In the remainder of this section, the acquisition setup and the statistical analysis of the acquired \gls{vr} traffic traces will be presented.

\subsection{Acquisition Setup} % (fold)
\label{sub:acquisition_setup}

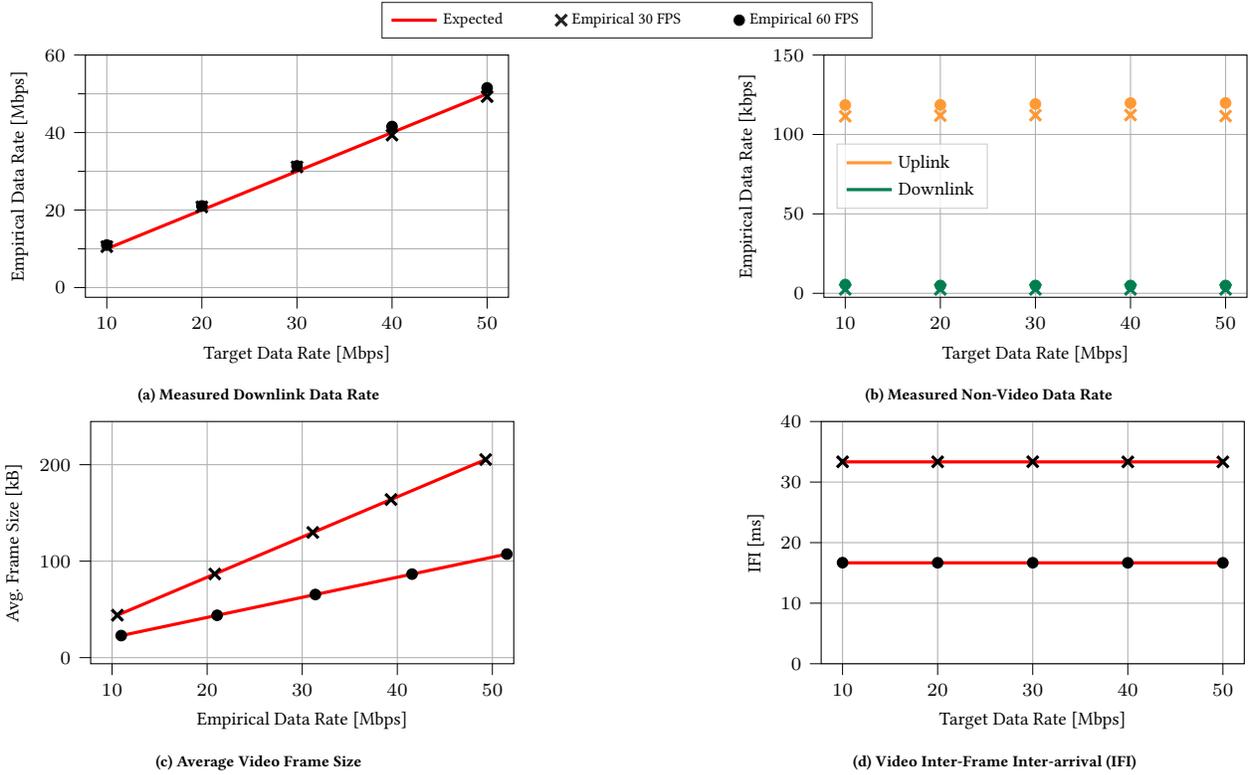
\begin{figure*}[t!]
\newcommand\fheight{0.6\columnwidth}
\newcommand\fwidth{0.9\columnwidth}
\begin{subfigure}[t]{\textwidth}
  \centering
  \input{img/legend_acquisitions.tex}
 \end{subfigure}
\\
 % \hspace*{\fill}%
 \begin{subfigure}[t]{0.45\textwidth}
  \centering
  \input{img/data_rate.tex}
  \caption{Measured Downlink Data Rate}
  \label{fig:measured_rate_vs_data_rate}
 \end{subfigure}
 \hspace*{\fill}%
 \begin{subfigure}[t]{0.45\textwidth}
  \centering
  \input{img/non_video_traffic.tex}
  \caption{Measured Non-Video Data Rate}
  \label{fig:non_video_traffic_vs_data_rate}
 \end{subfigure}
 \hspace*{\fill}%
 \\
 \begin{subfigure}[t]{0.45\textwidth}
  \centering
  \input{img/frame_size.tex}
  \caption{Average Video Frame Size} 
  \label{fig:frame_size_vs_data_rate}
 \end{subfigure}
 \hspace*{\fill}%
 \begin{subfigure}[t]{0.45\textwidth}
  \centering
  \input{img/ifi.tex}
  \caption{Video \acrfull{ifi}}
  \label{fig:ifi_vs_data_rate}
 \end{subfigure}
 % \hspace*{\fill}%

 \caption{Results from Acquired \gls{vr} Traffic Traces}
 \label{fig:vr_traces_vs_rate}
\end{figure*}

The setup comprises a desktop PC (equipped with an i7 processor, 32~GB of RAM, and an nVidia GTX~2080Ti graphics card) acting as a rendering server, and transmitting the information to a smartphone acting as a passive VR headset.
The two nodes are connected via USB tethering to avoid random interference from other surrounding devices and the less stable wireless channel.

To stream the VR traffic, the rendering server runs the application \textit{RiftCat~2.0}, connected to the \textit{VRidge} app running on the smartphone.\footnote{riftcat.com/vridge}
This setup allows the user to play VR games on the SteamVR platform for up to 10~minutes, enough to obtain multiple traffic traces that can be analyzed later.
The application allows for a number of settings, most notably (i) the display resolution and scan format (kept fixed at the smartphone's native resolution, i.e., 1920$\times$1080p), (ii) the frame rate, allowing the user to choose between 30~FPS and 60~FPS, and (iii) the target data rate (i.e., the data rate the application will try to consistently stream to the client) which can be set from 1~to 50~Mbps.

To simplify the analysis of the traffic stream in this first work, no VR games were played, acquiring only traces from the SteamVR waiting room with no audio.
The waiting room is still rendered and streamed in real time, thus allowing the capture of an actual VR stream.
Furthermore, in order for SteamVR to fully start and load the waiting room, traces were trimmed down to approximately 550~s each.

We noticed that the encoder used to stream the video from the rendering server to the phone was able to reduce the stream size when the scene was fairly static, thus preventing us from reaching data rates higher than 20~Mbps with the phone in a fixed position.
Small movements from the phone were sufficient to obtain a data rate close to the target one.

Traffic traces were obtained using Wireshark, a popular open-source packet analyzer, running on the rendering server and sniffing the tethered USB connection.
The traffic analysis was performed at 30~and 60~FPS for target data rates of \{10, 20, 30, 40, 50\}~Mbps, for a total of over 90~minutes of analyzed \gls{vr} traffic.

% subsection acquisition_setup (end)

\subsection{Traffic Analysis} % (fold)
\label{sub:traffic_analysis}

By analyzing the sniffed packet traces, we discovered that \textit{VRidge} uses UDP sockets over IPv4 and that the \gls{ul} stream contains several types of packets, such as synchronization, video frame reception information, and frequent small head-tracking information packets.
In \gls{dl}, instead, we found synchronization, acknowledgment, and video frame packets bursts.
We also found out that the application stream is based on ENet, a simple and robust network communication layer on top of UDP.

Video traffic is, as expected, the main source of data transmission (\cref{fig:measured_rate_vs_data_rate}).
Video frames are easily categorized by their transmission pattern: a single frame is fragmented into multiple smaller 1278~B packets sent back-to-back.
By reverse-engineering the bits composing the UDP payload, it was possible to identify 5 ranges of information in what appears to be a 31~B APP-layer header, specifically (i) the frame sequence number, (ii) the number of fragments composing the frame, (iii) the fragment sequence number, (iv) the total frame size, and (v) a checksum.
Thanks to this, we were able to reliably gather information on video frames, allowing for robust data processing.

\cref{fig:non_video_traffic_vs_data_rate} shows that \Gls{ul} traffic only accounts for 110~kbps for 30~FPS acquisitions and 120~kbps for 60~FPS acquisitions, while non-video \gls{dl} traffic is only about 2.5~kbps for 30~FPS acquisition and 5~kbps for 60~FPS acquisitions, regardless of the target data rate, and for this reason, they were ignored in the proposed analysis.

It follows that, considering $R$ the target data rate and $F$ the application frame rate, the \textit{average video frame size} is expected to be close to the ideal $S=R/F$, as shown in \cref{fig:frame_size_vs_data_rate}.
Note that the $x$-axis reports the empirical data rate rather than the target data rate, i.e., the average data rate estimated from the acquired traces, which differ slightly as shown in \cref{fig:measured_rate_vs_data_rate}.

Furthermore, \cref{fig:ifi_vs_data_rate} shows that the average \gls{ifi} perfectly matches the expected $1/F$.

Usually, when compressing a video, both intra-frame and inter-frame compression techniques are exploited.
Specifically, \glspl{iframe} use compression techniques similar to those of a simple static picture.
Instead, \glspl{pframe} exploit the temporal correlation of successive frames to greatly reduce the compressed frame size.
Similarly, \glspl{bframe} exploit the knowledge from subsequent frames as well, other than previous frames, to further improve the compression efficiency at the cost of non-real-time transmission.
Video compression standards such as H.264~\cite{h264Book} define the pattern of compressed frame types between two consecutive \glspl{iframe}, which is commonly referred to as \gls{gop}.

The traffic traces show that \glspl{gop} are not deterministic, and tend to be larger at lower target data rates, likely to take advantage of the higher compression generally provided by \glspl{pframe}.
Furthermore, lower target data rates may introduce a small delay to also encode the more efficient \glspl{bframe}, thus improving the visual quality while decreasing the overall responsiveness.
However, the strategy used by the application to map the specified target rate into a certain GoP format is proprietary and undisclosed, so that we could only observe some general trends.

% subsection traffic_analysis (end)

% section vr_traffic_acquisition_and_analysis (end)

\section{Traffic Model} % (fold)
\label{sec:traffic_model}

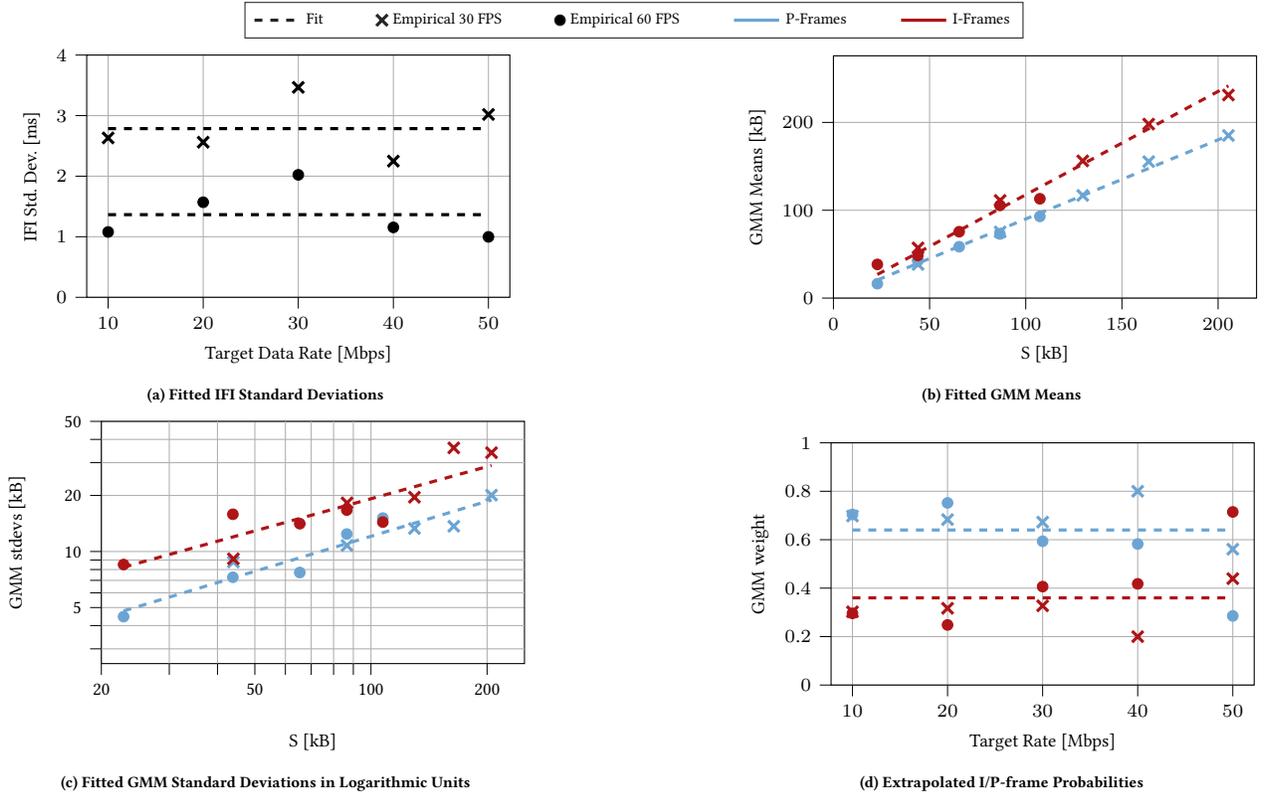
\begin{figure*}[t!]
\newcommand\fheight{0.6\columnwidth}
\newcommand\fwidth{0.9\columnwidth}

\begin{subfigure}[t]{\textwidth}
  \centering
  \input{img/legend_fit.tex}
 \end{subfigure}
\\
  % \hspace*{\fill}%
  \begin{subfigure}[t]{0.45\textwidth}
    \centering
    \input{img/ifi_stdev.tex}
    \caption{Fitted IFI Standard Deviations}
    \label{fig:ifi_stdev}
  \end{subfigure}
  \hspace*{\fill}%
  \begin{subfigure}[t]{0.45\textwidth}
    \centering
    \input{img/gmm_means_S.tex}
    \caption{Fitted GMM Means}
    \label{fig:gmm_means}
  \end{subfigure}
  \\
  % \hspace*{\fill}%
  \begin{subfigure}[t]{0.45\textwidth}
    \centering
    \input{img/gmm_stdevs_S.tex}
    \caption{Fitted GMM Standard Deviations in Logarithmic Units}
    \label{fig:gmm_stdevs}
  \end{subfigure}
  \hspace*{\fill}%
  \begin{subfigure}[t]{0.45\textwidth}
    \centering
    \input{img/gmm_probs.tex}
    \caption{Extrapolated I/P-frame Probabilities}
    \label{fig:gmm_probs}
  \end{subfigure}

  \caption{Fitted Models. For GMMs, the Red Lines Correspond to \gls{iframe} Statistics, the Blue Ones Correspond to \gls{pframe} Statistics}
  \label{fig:param_fit}
\end{figure*}

In this section, we will describe how we model frame sizes and inter-frame periodicity, leaving the model validation to \cref{sec:model_validation_and_possible_use_cases}.

\subsection{Modeling Frame Periodicity} % (fold)
\label{sub:modeling_frame_periodicity}

To fully characterize and thus generate a realistic frame period, more information is needed such as (i) the distribution of the frame period, (ii) the parameters of this distribution, (iii) the correlation between successive frame periods, and (iv) the correlation between the current frame size and the frame period.

To simplify the model, in this first analysis we assume the current frame size and the frame period to be independent, and we consider the stochastic process representing the frame period to be uncorrelated.
We thus focus only on the distribution of the frame period.

Among all measurements taken at both 30~and 60~FPS, the logistic distribution was often the best-fitting one, among all of the tested distributions.
For this reason, we choose to model the frame periods as independent and identically Logistic-distributed random variables $X\sim \mathrm{Logistic}(\mu,s)$ with \gls{pdf}
\begin{equation}\label{eq:logistic_distrib}
  p(x | \mu, s) = \frac{e^{-(x-\mu)/s}}{s(1+e^{-(x-\mu)/s})^2},
\end{equation}
where $\mu$ is the \textit{location} parameter, $s>0$ is the \textit{scale} parameter, and $\mathbb{E}[X]=\mu$, $\mathrm{std}(X) = \frac{s\pi}{\sqrt{3}}$.

Given the great accordance between the expected frame period and the empirical one (shown in \cref{fig:ifi_vs_data_rate}), we can easily set $\mathbb{E}[X]=\mu=\frac{1}{F}$.
Instead, to model the standard deviation of the proposed Logistic random variables, we need to further process the acquired data, although for a powerful enough rendering server we expect the standard deviation to also be inversely proportional to the frame rate.

\cref{fig:ifi_stdev} shows the average standard deviation of the acquired traces at both 30~and 60~FPS.
We notice that the average standard deviation at 30~FPS is 2.7855~ms, while at 60~FPS it is equal to 1.3646~ms, with a perfect ratio of 2.04.
This suggests that our rendering server is powerful enough to reliably handle streams at both frame rates since the standard deviation of the \gls{ifi} nearly doubles as the frame rate halves, as expected.

While the data is not enough for a robust generalization, it does suggest that a higher frame rate $F$ yields lower \gls{ifi} standard deviation $d$, keeping a roughly constant ratio with the average \gls{ifi}, with an inverse law:  $d = \frac{c}{F}$.
Assuming that this inverse law holds, we compute the average $c$ obtained by the sets of acquisitions at 30~and 60~FPS, equal to $c=0.0827$.

% subsection modeling_frame_periodicity (end)

\subsection{Modeling Frame Size} % (fold)
\label{sub:modeling_frame_sizes}

Following the discussion in \cref{sub:traffic_analysis}, we propose to model the frame size distribution of the video frame with a \gls{gmm}, i.e., $V(S)\sim GMM(\bm{\mu}(S),\bm{\sigma}^2(S))$, with \gls{pdf}
\begin{equation}\label{eq:gmm}
  V(S) = \chi_I(S) V_I(S) + (1-\chi_I(S)) V_P(S),
\end{equation}
where $\chi_I(S)$ is the indicator function for \glspl{iframe} which takes the value of 1 with probability $w_I(S)$ and 0 otherwise, and $V_f(S)\sim \mathcal{N} \left( \mu_f(S), \sigma_f^2(S) \right)$, $f\in\{I, P\}$.
Clearly, the fitted normal variable with the lower mean will be associated to \glspl{pframe} while the one with the higher mean will be associated to \glspl{iframe}.

To generalize the model, the parameters of the GMM should be extended to arbitrary target data rates.%\ml{ and frame rates}.

Starting from the GMMs of the acquired traffic traces, the mean frame sizes of I- and \glspl{pframe} are generalized by fitting linear models, while their standard deviations are better fitted by a power law, both as a function of the expected average frame size $S$.
Since a target data rate approaching zero would require video frames to also approach zero, we force the linear fit to have no intercept, i.e., $\mu_I(S) = s_I S$, $\mu_P(S) = s_P S$, $\sigma_I(S) = a_I S^{b_I}$, and $\sigma_P(S) = a_P S^{b_P}$ as depicted with dashed lines in \cref{fig:gmm_means,fig:gmm_stdevs}.

By setting $\mathbb{E}[V(S)]$ equal to $S$ we get
\begin{equation}
\begin{cases}
  \mathbb{E}[V(S)] = w_I(S) \mu_I(S) + w_P(S) \mu_P(S) = S\\
  w_I + w_P = 1
\end{cases},
\end{equation}
from which $w_I(S)=w_I=\frac{1 - s_P}{s_I - s_P}$ and $w_P(S)=w_P=\frac{s_I-1}{s_I - s_P}$, regardless of $S$.
Setting $0\leq w_I, w_P \leq 1$ results in a requirement for $s_I$ and $s_P$, specifically, $s_P \leq 1 \leq s_I$.

To make the model more robust, we first fit the GMM 50~times with random initial conditions and pick the best-fitting model, and then weigh the linear fit of the parameters proportionally to the goodness of the \gls{gmm} fit.

To improve the reliability of our model, when fitting the GMM means and standard deviations we use the empirical data rate as the independent variable instead of the target data rate since the two differ slightly as noticed in \cref{sub:traffic_analysis}.
In fact, we aim at modeling the general behavior of the application while trying to generate the amount of traffic requested by the user as closely as possible.

\cref{fig:gmm_means} shows that the means of the GMMs are fitted well by a linear model in the $S$ domain, which is used to simultaneously process the data of the acquisitions obtained both at 30~and 60~FPS, yielding the fitter parameters $s_I=1.1764$ and $s_P=0.9008$.
The fitted slopes result in $w_P=0.64$, which fits well the 30~FPS acquisitions, with slightly worse performance at 60~FPS.

On the other hand, the GMMs standard deviations show a more noisy fit (\cref{fig:gmm_stdevs}), yielding parameters $a_I=26.2065$, $b_I=0.5730$, $a_P=9.0399$, and $b_P=0.6251$.
This suggests an approximate square-root relationship between the average frame size $S$ and the standard deviation of the GMMs.
The logarithmic plot helps the visualization by transforming a power law into a simple linear relationship.

Frame sizes are independently drawn from the mixture model instead of simulating a \gls{gop}, since different \glspl{gop} were found for different target data rates, and always with a non-deterministic nature.

% subsection modeling_frame_sizes (end)

% section traffic_model (end)

\section{Ns-3 Implementation} % (fold)
\label{sec:ns_3_implementation}

\glsunset{ns3}

To properly model and test the performance of VR traffic over a simulated network, a flexible application framework has been implemented in \gls{ns3} and made publicly available~\cite{ns-3-vr-app}.
The framework is based on the ns-3.33 release and aims at providing a novel additional traffic model, easily customizable by the final user.

The proposed framework allows the user to send packet bursts fragmented into multiple packets by \texttt{Bursty\-Application}, later re-aggregated at the receiver, if possible, by \texttt{Burst\-Sink}.
Since the generation of packet bursts is crucial to model a wide range of possibilities, a generic \texttt{Burst\-Generator} interface has been defined.
Users can implement arbitrary generators by extending this interface, and three examples have been provided and will be described in \cref{sub:burst_generator_interface}.
Finally, each fragment comprises a novel \texttt{Seq\-Ts\-Size\-Frag\-Header}, which includes information on both the fragment and the current burst, allowing \texttt{Burst\-Sink} to correctly re-aggregate or discard a burst, yielding information on received fragments, received bursts, and failed bursts.

More details on the implementation and the rationale behind these applications will be given in the following sections.

\subsection{Bursty Application} % (fold)
\label{sub:bursty_application}

Inspired by the acquired traffic traces described in \cref{sub:acquisition_setup}, \texttt{Bursty\-Application} periodically sends bursts of data divided into multiple smaller fragments of (at most) a given size.
Since burst size and period statistics can be quite general, the generation of the burst statistics is delegated to objects extending the \texttt{Burst\-Generator} interface, later described in \cref{sub:burst_generator_interface}.
\texttt{Bursty\-Helper} is also implemented to simplify the generation and installation of \texttt{Bursty\-Application}s with given \texttt{Burst\-Generator}s to network nodes and examples are provided.

Each fragment carries a \texttt{Seq\-Ts\-Size\-Frag\-Header}, an extension of \texttt{Seq\-Ts\-Size\-Header} which adds the information on the fragment sequence number and the total number of fragments composing the burst, on top of the (burst) sequence number and size as well as the transmission time-stamp.
After setting a desired \texttt{Fragment\-Size} in bytes, the application will compute how many fragments will be generated to send the full burst to the target receiver, although the last two fragments may be smaller due to the size of the burst not being a multiple of the fragment size, and the presence of the extra header.

Traces notify the user when fragments and bursts are sent, while also keeping track of the number of bursts, fragments, and bytes sent, making it easier to quickly compute some simple high-level metrics directly from the main script of the simulation.

% subsection bursty_application (end)

\subsection{Burst Generator Interface} % (fold)
\label{sub:burst_generator_interface}

A generic bursty application can show extremely different behaviors.
For example, an application could send a given amount of data periodically in a deterministic fashion, or the burst size or the period could be random with arbitrary statistics, successive bursts could be correlated (e.g., the concept of \gls{gop} for video-coding standards such as H.264~\cite{h264Book}), and even the burst size and the time before the next burst might be correlated.

To accommodate for the widest range of possibilities, a \texttt{Burst\-Generator} interface has been defined.
Classes extending this interface must define two pure virtual functions:
\begin{enumerate}
  \item \texttt{Has\-Next\-Burst}: to ensure that the burst generator is able to generate a new burst size and the time before the next burst (also called \textit{next period} in the remainder of this paper);
  \item \texttt{Generate\-Burst}: yielding the burst size of the current burst as well as the next period, if it exists.
\end{enumerate}

Three classes extending this interface are proposed and briefly discussed in the remainder of this section, allowing users to generate very diverse statistics without the need to implement their own custom generator in most cases.

\paragraph{Simple Burst Generator} % (fold)
\label{par:simple_burst_generator}

Inspired from \texttt{On\-Off\-Application}, \texttt{Sim\-ple\-Burst\-Generator} defines the current burst size and the next period as generic \texttt{Random\-Variable\-Stream}s.
Users are thus able to model arbitrary burst size and next period distributions, by: using the distributions already implemented in ns-3; implementing more distributions; or simply defining arbitrary \glspl{cdf} for \texttt{Empirical\-Random\-Variable}s.

Limitations for this generator lie in the correlation of the generated random variables: burst size and next period are independently drawn as are successive bursts.

% paragraph simple_burst_generator (end)
\begin{figure*}[t!]
  \newcommand\fheight{0.6\columnwidth}
  \newcommand\fwidth{0.9\columnwidth}

   \begin{subfigure}[t]{\textwidth}
    \centering
    \input{img/cdfs_ifi_legend.tex}
   \end{subfigure}
   \\
   % \hspace*{\fill}%
   \begin{subfigure}[t]{0.45\textwidth}
    \centering
    \input{img/ecdf_ifi_30fps.tex}
    \caption{30~FPS}
    \label{fig:ecdf_ifi_30fps}
   \end{subfigure}
   \hspace*{\fill}%
   \begin{subfigure}[t]{0.45\textwidth}
    \centering
    \input{img/ecdf_ifi_60fps.tex}
    \caption{60~FPS}
    \label{fig:ecdf_ifi_60fps}
   \end{subfigure}
   % \hspace*{\fill}%

 \caption{Comparison of Empirical CDFs of \acrlongpl{ifi}}
 \label{fig:ecdfs_ifi}
\end{figure*}

\paragraph{VR Burst Generator} % (fold)
\label{par:vr_burst_generator}

\texttt{Vr\-Burst\-Generator} is a direct implementation of the model proposed in \cref{sec:traffic_model}, where bursts model video frames.

Similar to the \textit{RiftCat} software described in \cref{sub:acquisition_setup}, this generator makes it possible to choose a target data rate and a frame rate.

While traces were taken at specific frame rates and target data rates, the proposed model attempts to generalize them, although without any knowledge on the quality of the generalization beyond the boundaries imposed by the streaming software.

To generate the frame size and the next period, \texttt{Logistic\-Random\-Variable} and \texttt{Mixture\-Random\-Variable} have been implemented in ns-3.

A validation of the proposed model based on this burst generator will be discussed in \cref{sec:model_validation_and_possible_use_cases}.

% paragraph vr_burst_generator (end)

\paragraph{Trace File Burst Generator} % (fold)
\label{par:trace_file_burst_generator}

Finally, users might want to reproduce in ns-3 a traffic trace obtained by a real application, generated by a separate traffic generator, or even manually written by a user (e.g., for static debugging/testing purposes).
For these reasons, \texttt{Trace\-File\-Burst\-Generator} was introduced, taking advantage of \texttt{Csv\-Reader} to parse a csv-like file declaring a (burst size, next period) pair for each row.
Once traces are imported, the generator will sequentially yield every burst, returning \texttt{false} as output to \texttt{TraceFileBurstGenerator::HasNextBurst} after the last row of the trace file is yielded, thus stopping the \texttt{Bursty\-Application}.

A \texttt{StartTime} can be set as an attribute, allowing the user to control which part of the file trace will be used in the simulation.
This can be especially useful when the total simulation duration is shorter than the traffic trace, making it possible to decouple users by setting different start times.

Several VR traffic traces using different frame rates and target data rates are available~\cite{ns-3-vr-app} in the described format for a total of over 90~minutes of processed acquisitions, comprising some relevant metadata as part of the commented header.
Interested readers can thus simulate real VR video traffic in their ns-3 simulations, or expand the analysis performed in \cref{sub:traffic_analysis,sec:traffic_model}.

% paragraph trace_file_burst_generator (end)

% subsection burst_generator_interface (end)

\subsection{Burst Sink} % (fold)
\label{sub:burst_sink}

\begin{figure*}[t!]
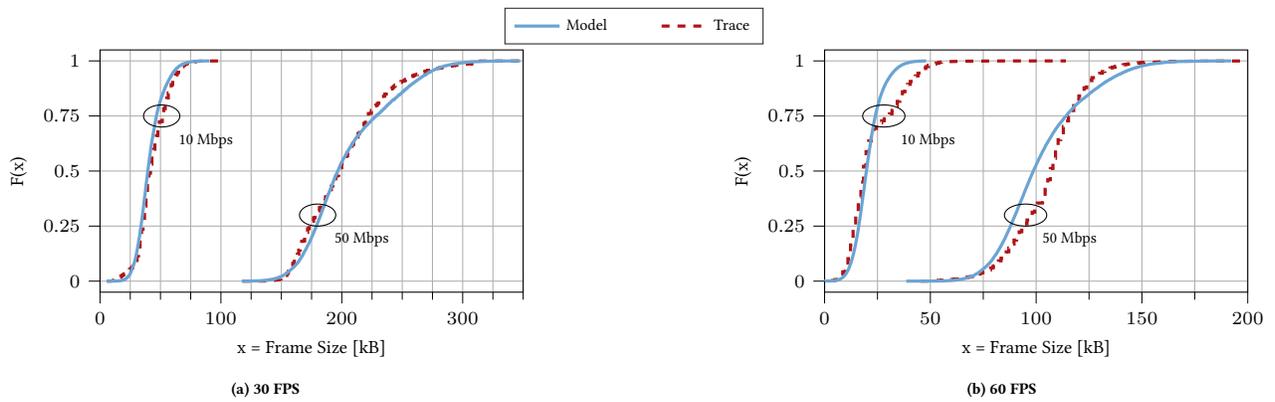

  \newcommand\fheight{0.6\columnwidth}
  \newcommand\fwidth{0.9\columnwidth}

   \begin{subfigure}[t]{\textwidth}
    \centering
    \input{img/cdfs_frame_legend.tex}
   \end{subfigure}
   \\
 % \hspace*{\fill}%
 \begin{subfigure}[t]{0.45\textwidth}
  \centering
  \input{img/ecdf_frame_30fps.tex}
  \caption{30~FPS}
  \label{fig:ecdf_frame_30fps}
 \end{subfigure}
 \hspace*{\fill}%
 \begin{subfigure}[t]{0.45\textwidth}
  \centering
  \input{img/ecdf_frame_60fps.tex}
  \caption{60~FPS}
  \label{fig:ecdf_frame_60fps}
 \end{subfigure}
 % \hspace*{\fill}%

 \caption{Comparison of Empirical CDFs of Frame Sizes}
 \label{fig:ecdfs_frame}
\end{figure*}

An adaptation of the existing \texttt{Packet\-Sink}, called \texttt{Burst\-Sink}, is proposed for the developed bursty framework.
This new application expects to receive packets from users equipped with \texttt{Bursty\-Application}s and tries to re-aggregate fragments into packets.

While the current version of \texttt{Packet\-Sink} is able to assemble byte streams with \texttt{Seq\-Ts\-Size\-Header}, there are two reasons why \texttt{Burst\-Sink} was created, specifically (i) to stress the dependence of this framework on UDP rather than TCP sockets, as the acquisitions suggested, thus expecting individual fragments sent unreliably rather than a reliable byte stream, and (ii) to trace the reception at both the fragment and the burst level.

The application implements a simple best-effort aggregation algorithm, assuming that (i) the burst transmission duration is much shorter than the \textit{next period}, and (ii) all fragments are needed to re-aggregate a burst.
Specifically, fragments of a given burst are collected, even if unordered, and, if all fragments are received, the burst is successfully received.
If, instead, fragments of subsequent bursts are received before all fragments of the previous one, then the previous burst is discarded.
Information on the current fragment and burst can easily be recovered from \texttt{Seq\-Ts\-Size\-Frag\-Header}, allowing the application to verify whether a burst has been fully received or not.
If needed and suggested by real-world applications, future works might also introduce the concept of APP-level \gls{fec}.

Traces notify the user when fragments are received and when bursts are successfully received or discarded, together with all the related relevant information.
Furthermore, similarly to the \texttt{Bursty\-Application}, also the \texttt{Burst\-Sink} application keeps track of the number of bursts, fragments, and bytes received.

% subsection burst_sink (end)

% section ns_3_implementation (end)

\section{Model Validation and Possible Use Cases} % (fold)
\label{sec:model_validation_and_possible_use_cases}

  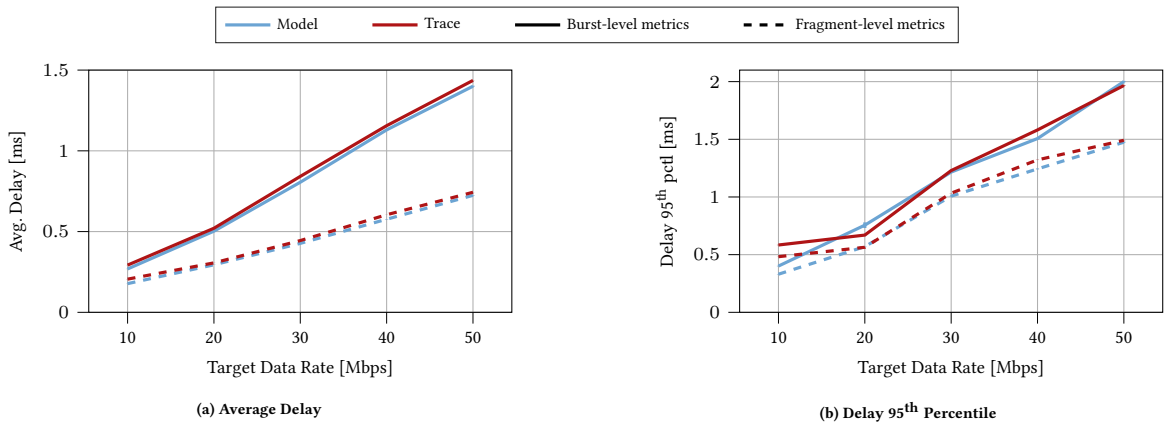
\begin{figure*}[t!]
  \newcommand\fheight{0.6\columnwidth}
  \newcommand\fwidth{0.9\columnwidth}
   
    \begin{subfigure}[t]{\textwidth}
      \centering
      \input{img/appRate/legend.tex}
    \end{subfigure}
    \\
    \vspace*{1ex}
    \hspace*{\fill}%
    \begin{subfigure}[t]{0.45\textwidth}
      \centering
      \input{img/appRate/delay_avg.tex}
      \caption{Average Delay}
      \label{fig:avg_delay_apprate}
    \end{subfigure}
    \hspace*{\fill}%
    \begin{subfigure}[t]{0.45\textwidth}
      \centering
      \input{img/appRate/delay_95.tex}
      \caption{Delay 95\textsuperscript{th} Percentile}
      \label{fig:delay_95_apprate}
    \end{subfigure}
    \hspace*{\fill}
  
    \caption{APP-layer Metrics Plotted Against an Increasing Target Data Rate with 60~FPS Sources}
    \label{fig:apprate}
  \end{figure*}

  \begin{figure*}[t!]
  \newcommand\fheight{0.6\columnwidth}
  \newcommand\fwidth{0.9\columnwidth}
   
    \begin{subfigure}[t]{\textwidth}
      \centering
      \input{img/nStas/legend.tex}
    \end{subfigure}
    \\
    \vspace*{1ex}
    \hspace*{\fill}%
    \begin{subfigure}[t]{0.45\textwidth}
      \centering
      \input{img/nStas/delay_avg.tex}
      \caption{Average Delay}
      \label{fig:avg_delay_nstas}
    \end{subfigure}
    \hspace*{\fill}%
    \begin{subfigure}[t]{0.45\textwidth}
      \centering
      \input{img/nStas/delay_95.tex}
      \caption{Delay 95\textsuperscript{th} Percentile}
      \label{fig:delay_95_nstas}
    \end{subfigure}
    \hspace*{\fill}%
  
    \caption{APP-layer Metrics Plotted Against an Increasing Number of STAs.
    Each STA Uses a VR Application with a Target Data Rate of 50~Mbps
    }
    \label{fig:nstas}
  \end{figure*}
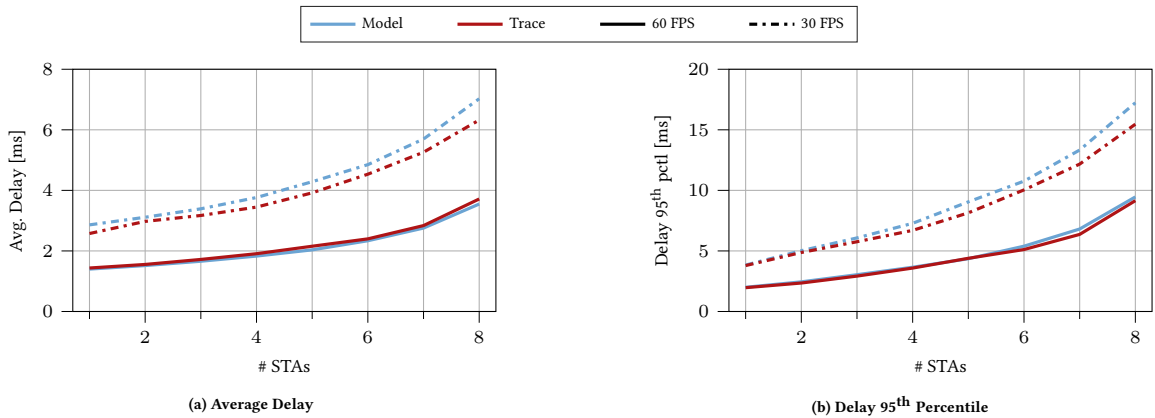

This section will present a comparison between the acquired VR traces and the proposed model, as well as an example use case.

For both the comparison and the example, we show the results of full-stack simulations highlighting the importance of accurately modeling a traffic source by using (i) the proposed model and (ii) the acquired traffic traces.
For full-stack simulations we consider a simple \gls{wifi} network based on IEEE~802.11ac, sending data over a single stream and using MCS~9 over a 160~MHz channel.

\subsection{Model Validation} % (fold)
\label{sub:model_validation}

A comparison between the modeled distributions and the acquired traffic traces is shown in \cref{fig:ecdfs_ifi,fig:ecdfs_frame}.

In particular, as expected from \cref{fig:ifi_stdev}, the \gls{ifi} standard deviation is a loose fit, and thus \glspl{cdf} shown in \cref{fig:ecdfs_ifi} show a fairly large deviation between the model and two examples of acquired traces, namely the highest and lowest target data rates acquired for both available frame rates.
This is to be expected given the large variance of the acquired data, although the objective of our model is to generalize the behavior of a real application for both data rates and frame rates, making it extremely hard to represent well the acquired data.

\cref{fig:ecdfs_frame} shows the \glspl{cdf} for frame sizes at different frame rate and target data rate.
The 30~FPS data is overall fitted well by the proposed model, and two examples at different target data rates are shown.
On the other hand, the 60~FPS data shows a different behavior with respect to the model.
A first explanation is that the 60~FPS traces always have a higher empirical data rate than the target one, so the red dashed lines are generally to the right of the blue lines.
Furthermore, doubling the frame rate at a constant data rate halves the average video frame size, making it necessary to use more sophisticated encoding techniques to still be able to obtain a good enough video quality by biasing the relative size and frequency of I-~and P-frames.
Still, our model is able to capture the overall distributions, making it possible to generate novel traffic data with parameters that were never acquired.

End-to-end results in \cref{fig:apprate} show good accordance between models and empirical data, except for the 95\textsuperscript{th} percentile of the delay.
In fact, the proposed model shown in blue shows good accordance with the traffic traces, suggesting that it is able to emulate sufficiently well the traffic statistics to obtain similar full-stack results.

It is also important to notice the difference between fragment-wise and burst-wise statistics.
For applications such as VR, where the whole burst needs to be received to acquire the desired information, it is crucial to measure burst-level performance to get a deeper understanding of the system performance.
In general, fragment-level metrics will be more optimistic than those at the burst level, which may lead to incorrect conclusions.

% subsection model_validation (end)

\subsection{Examples of Use Cases} % (fold)
\label{sub:examples_of_use_cases}

To exemplify the uses of the proposed model, we discuss a possible scenario of interest where we test how well an IEEE~802.11ac network can support intense \gls{vr} traffic, for example in a VR arena.

In \cref{fig:nstas}, we show the simulation results for a scenario with multiple users running VR applications with a target rate of 50~Mbps in a \gls{wifi} network.
We compare the acquired trace file, where \glspl{sta} import and generate disjoint parts of the trace file, with the proposed model, both at 30~and 60~FPS.

Notice that the fixed target data rate of $R=$50~Mbps, average video frame size $S$ and frame rate $F$ will have the same ratio resulting in double the frame size for 30~FPS streams with respect to 60~FPS.
For a network with fixed channel capacity, this translates to a delay which will also double assuming that processing, queuing, and other delays independent of the burst size are negligible, hence explaining the different delay performance of the two frame rates in \cref{fig:nstas}.

From \cref{fig:avg_delay_nstas}, it is possible to see that the average burst delay is below the maximum tolerable delay of 5-9~ms specified in~\cite{huaweiVrArWhitePaper} for up to 8~users at both 30~and 60~FPS, although the delays of the 30~FPS streams are higher than those of the 60~FPS streams, as expected.

Instead, if a good overall quality of experience should be granted, the same bound for the 95\textsuperscript{th} percentile of the delay would only allow up to 5~users in the system for 30~FPS streams, or 7~users for 60~FPS streams, as shown in \cref{fig:delay_95_nstas}, thus greatly reducing the arena capacity for increased reliability and overall user experience.

% subsection examples_of_use_cases (end)

% section model_validation_and_possible_use_cases (end)

\section{Conclusions} % (fold)
\label{sec:conclusions}

In this paper, we presented a simple \gls{vr} traffic model based on over 90~minutes of acquired traffic traces.
While being simple, ignoring second-order statistics, and being based on an ideal setting, this model marks a starting point for network analysis and optimization tailored for this novel and peculiar type of traffic, introducing a more realistic traffic model into ns-3.

The proposed ns-3 framework for bursty applications is publicly available and open source~\cite{ns-3-vr-app}, together with the implementation of the proposed traffic model and the actual traffic traces experimentally obtained.
We also attempted to generalize the model to arbitrary target data rates and frame rates, allowing users to experiment with arbitrary application-level settings that suit their specific research.

The model has been built upon a framework to simulate bursty applications in ns-3, where burst size and period can be customized with little additional code, and traces for burst-level metrics collections allow the user to better analyze a complex application QoS.

Future works will focus on improving the quality and generality of this approach.
For example, second-order statistics will be taken into account, trying to better characterize the statistics of \glspl{gop}.
More acquisitions will be taken, possibly longer, with different streaming and video encoding settings, on several \gls{vr} applications.
Finally, more complex settings will be considered, e.g., adding head movements, in order to analyze possible correlations between them and the generated traffic.

% section conclusions (end)

\begin{acks}
Mattia Lecci's activities were supported by \textit{Fondazione CaRiPaRo} under the grant ``Dottorati di Ricerca 2018.''
This work was partially supported by NIST under Award No.~60NANB19D122.
\end{acks}

\bibliographystyle{ACM-Reference-Format}
\bibliography{bibl}

\end{document}

%% file: acronyms.tex
\newacronym{3gpp}{3GPP}{3rd Generation Partnership Project}
\newacronym{5g}{5G}{5\textsuperscript{th} Generation}
\newacronym{5gc}{5GC}{5G Core}
\newacronym{bs}{BS}{Base Station}
\newacronym{abft}{A-BFT}{Association-BeamForming Training}
\newacronym[firstplural=Access Categories (ACs)]{ac}{AC}{Access Category}
\newacronym{adc}{ADC}{Analog to Digital Converter}
\newacronym{addts}{ADDTS}{Add Traffic Stream}
\newacronym{afbw}{AFBW}{Average Fading Bandwidth}
\newacronym{aid}{AID}{Association ID}
\newacronym{aimd}{AIMD}{Additive Increase Multiplicative Decrease}
\newacronym{am}{AM}{Acknowledged Mode}
\newacronym{amc}{AMC}{Adaptive Modulation and Coding}
\newacronym{ampdu}{A-MPDU}{MAC Protocol Data Unit Aggregation}
\newacronym{aoa}{AoA}{Angle of Arrival}
\newacronym{aod}{AoD}{Angle of Departure}
\newacronym{ap}{AP}{Access Point}
\newacronym{app}{APP}{Application Layer}
\newacronym{aqm}{AQM}{Active Queue Management}
\newacronym{ar}{AR}{Augmented Reality}
\newacronym{arf}{ARF}{Auto Rate Fallback}
\newacronym{arp}{ARP}{Address Resolution Protocol}
\newacronym{ati}{ATI}{Announcement Transmission Interval}
\newacronym{awgn}{AGWN}{Additive White Gaussian Noise}
\newacronym{awv}{AWV}{Antenna Weight Vector}
\newacronym{balia}{BALIA}{Balanced Link Adaptation}
\newacronym{bdp}{BDP}{Bandwidth-Delay Product}
\newacronym{ber}{BER}{Bit Error Rate}
\newacronym{bf}{BF}{Beamforming}
\newacronym{bframe}{B-frame}{Bipredictive-coded frame}
\newacronym{bhi}{BHI}{Beacon Header Interval}
\newacronym{bi}{BI}{Beacon Interval}
\newacronym{brp}{BRP}{Beam Refinement Protocol}
\newacronym{bss}{BSS}{Basic Service Set}
\newacronym{bti}{BTI}{Beacon Transmission Interval}
\newacronym{cad}{CAD}{Computer-aided Design}
\newacronym{cbap}{CBAP}{Contention-Based Access Period}
\newacronym{cbr}{CBR}{Constant Bitrate}
\newacronym{cc}{CC}{Congestion Control}
\newacronym{cdf}{CDF}{Cumulative Distribution Function}
\newacronym{cir}{CIR}{Channel Impulse Response}
\newacronym{cn}{CN}{Core Network}
\newacronym{cp}{CP}{Control Plane}
\newacronym{cqi}{CQI}{Channel Quality Indicator}
\newacronym{crs}{CRS}{Cell Reference Signal}
\newacronym{csirs}{CSI-RS}{Channel State Information - Reference Signal}
\newacronym{csmaca}{CSMA/CA}{Carrier Sense Multiple Access with Collision Avoidance}
\newacronym{cts}{CTS}{Clear to Send}
\newacronym{dc}{DC}{Dual Connectivity}
\newacronym{dce}{DCE}{Direct Code Execution}
\newacronym{dcf}{DCF}{Distributed Coordination Function}
\newacronym{dci}{DCI}{Downlink Control Information}
\newacronym{delts}{DELTS}{Delete Traffic Stream}
\newacronym{dl}{DL}{Downlink}
\newacronym{dmg}{DMG}{Directional Multi-Gigabit}
\newacronym{dmr}{DMR}{Deadline Miss Ratio}
\newacronym{dmrs}{DMRS}{DeModulation Reference Signal}
\newacronym{dti}{DTI}{Data Transmission Interval}
\newacronym{e2e}{E2E}{End-to-End}
\newacronym{ecn}{ECN}{Explicit Congestion Notification}
\newacronym{edca}{EDCA}{Enhanced Distributed Channel Access}
\newacronym{edf}{EDF}{Earliest Deadline First}
\newacronym{embb}{eMBB}{Enhanced Mobile BroadBand}
\newacronym{enb}{eNB}{evolved Node Base}
\newacronym{endc}{EN-DC}{E-UTRAN-\gls{nr} \gls{dc}}
\newacronym{epc}{EPC}{Evolved Packet Core}
\newacronym{es}{ES}{Edge Server}
\newacronym{ese}{ESE}{Extended Schedule Element}
\newacronym{fdd}{FDD}{Frequency Division Duplexing}
\newacronym{fdma}{FDMA}{Frequency Division Multiple Access}
\newacronym{fec}{FEC}{Forward Error Correction}
\newacronym{fov}{FoV}{Field-of-View}
\newacronym{fs}{FS}{Fast Switching}
\newacronym{ftp}{FTP}{File Transfer Protocol}
\newacronym{fr2}{FR2}{Frequency Range 2}
\newacronym{gmm}{GMM}{Gaussian Mixture Model}
\newacronym{gnb}{gNB}{Next Generation Node Base}
\newacronym[firstplural=Group of Pictures (GoPs)]{gop}{GoP}{Group of Pictures}
\newacronym{harq}{HARQ}{Hybrid Automatic Repeat reQuest}
\newacronym{hetnet}{HetNet}{Heterogeneous Network}
\newacronym{hh}{HH}{Hard Handover}
\newacronym{hol}{HOL}{Head-of-Line}
\newacronym{hqf}{HQF}{Highest-quality-first}
\newacronym{ia}{IA}{Initial Access}
\newacronym{iab}{IAB}{Integrated Access and Backhaul}
\newacronym{ibss}{IBSS}{Independent Basic Service Set}
\newacronym{id}{ID}{Identifier}
\newacronym{ifi}{IFI}{Inter-Frame Inter-arrival}
\newacronym{iframe}{I-frame}{Intra-coded frame}
\newacronym{imt}{IMT}{International Mobile Telecommunication}
\newacronym{imt2020}{IMT-2020}{International Mobile Te\-le\-com\-mu\-ni\-ca\-tion-2020}
\newacronym{inr}{INR}{Interference to Noise Ratio}
\newacronym{iot}{IoT}{Internet of Things}
\newacronym{ipa}{IPA}{Inter-Packet Arrival}
\newacronym{ism}{ISM}{Industrial, Scientific, and Medical}
\newacronym{itu}{ITU}{International Telecommunication Union}
\newacronym{ks}{KS}{Kolmogorov-Smirnov}
\newacronym{kpi}{KPI}{Key Performance Indicator}
\newacronym{lcf}{LCF}{Level Crossing Frequency}
\newacronym{lcm}{lcm}{least common multiple}
\newacronym{lcr}{LCR}{Level Crossing Rate}
\newacronym{los}{LoS}{Line-of-Sight}
\newacronym{lp}{LP}{Low Power}
\newacronym{lte}{LTE}{Long Term Evolution}
\newacronym{m2m}{M2M}{Machine to Machine}
\newacronym{mac}{MAC}{Medium Access Control}
\newacronym{mc}{MC}{Multi-Connectivity}
\newacronym{mcs}{MCS}{Modulation and Coding Scheme}
\newacronym{mec}{MEC}{Mobile Edge Cloud}
\newacronym{mi}{MI}{Mutual Information}
\newacronym{mib}{MIB}{Master Information Block}
\newacronym{mimo}{MIMO}{Multiple Input, Multiple Output}
\newacronym{mumimo}{MU-MIMO}{Multi-User Multiple Input, Multiple Output}
\newacronym{ml}{ML}{Machine Learning}
\newacronym{mlr}{MLR}{Maximum-local-rate}
\newacronym[plural=\gls{mme}s,firstplural=Mobility Management Entities (MMEs)]{mme}{MME}{Mobility Management Entity}
\newacronym{mmw}{mmW}{Millimeter Wave}
\newacronym{moi}{MoI}{Method of Images}
\newacronym{mpc}{MPC}{Multi Path Component}
\newacronym{mpdu}{MPDU}{MAC Protocol Data Unit}
\newacronym{mptcp}{MPTCP}{Multipath TCP}
\newacronym{mr}{MR}{Maximum Rate}
\newacronym{mrdc}{MR-DC}{Multi \gls{rat} \gls{dc}}
\newacronym{msdu}{MSDU}{MAC Service Data Unit}
\newacronym{mss}{MSS}{Maximum Segment Size}
\newacronym{mtd}{MTD}{Machine-Type Device}
\newacronym{mtu}{MTU}{Maximum Transmission Unit}
\newacronym{nav}{NAV}{Network Allocation Vector}
\newacronym{ncbr}{NCBR}{Non-Constant Bitrate}
\newacronym{nfv}{NFV}{Network Function Virtualization}
\newacronym{nlos}{NLoS}{Non-Line-of-Sight}
\newacronym{nr}{NR}{New Radio}
\newacronym{nrmse}{NRMSE}{Normalized Root Mean Square Error}
\newacronym{ns3}{ns-3}{Network Simulator 3}
\newacronym{nsa}{NSA}{Non Stand Alone}
\newacronym{o2i}{O2I}{Outdoor-to-Indoor}
\newacronym{ofdm}{OFDM}{Orthogonal Frequency Division Multiplexing}
\newacronym{pa}{PA}{Position-aware}
\newacronym{pan}{PAN}{Personal Area Network}
\newacronym{pbch}{PBCH}{Physical Broadcast Channel}
\newacronym{pbss}{PBSS}{Personal Basic Service Set}
\newacronym{pcf}{PCF}{Point Coordinator Function}
\newacronym{pcp}{PCP}{\gls{pbss} Central Point}
\newacronym{pcpap}{PCP/AP}{\acrlong{pcp}/\acrlong{ap}}
\newacronym{pdcch}{PDCCH}{Physical Downlonk Control Channel}
\newacronym{pdcp}{PDCP}{Packet Data Convergence Protocol}
\newacronym{pdf}{PDF}{Probability Density Function}
\newacronym{pdsch}{PDSCH}{Physical Downlink Shared Channel}
\newacronym{per}{PER}{Packet Error Rate}
\newacronym{pdu}{PDU}{Packet Data Unit}
\newacronym{pf}{PF}{Proportional Fair}
\newacronym{pframe}{P-frame}{Predictive-coded frame}
\newacronym{pgw}{PGW}{Packet Gateway}
\newacronym{phy}{PHY}{Physical Layer}
\newacronym{ppdu}{PPDU}{PHY Protocol Data Unit}
\newacronym{ppp}{PPP}{Poisson Point Process}
\newacronym{prb}{PRB}{Physical Resource Block}
\newacronym{pss}{PSS}{Primary Synchronization Signal}
\newacronym{pucch}{PUCCH}{Physical Uplink Control Channel}
\newacronym{pusch}{PUSCH}{Physical Uplink Shared Channel}
\newacronym{qd}{QD}{Quasi Deterministic}
\newacronym{qoe}{QoE}{Quality of Experience}
\newacronym{qos}{QoS}{Quality of Service}
\newacronym{rach}{RACH}{Random Access Channel}
\newacronym{ran}{RAN}{Radio Access Network}
\newacronym[firstplural=Radio Access Technologies (RATs)]{rat}{RAT}{Radio Access Technology}
\newacronym{red}{RED}{Random Early Detection}
\newacronym{rf}{RF}{Radio Frequency}
\newacronym{rl}{RL}{Reinforcement Learning}
\newacronym{rlc}{RLC}{Radio Link Control}
\newacronym{rlf}{RLF}{Radio Link Failure}
\newacronym{rr}{RR}{Round Robin}
\newacronym{rrc}{RRC}{Radio Resource Control}
\newacronym{rrm}{RRM}{Radio Resource Management}
\newacronym{rs}{RS}{Remote Server}
\newacronym{rsrp}{RSRP}{Reference Signal Received Power}
\newacronym{rsrq}{RSRQ}{Reference Signal Received Quality}
\newacronym{rss}{RSS}{Received Signal Strength}
\newacronym{rssi}{RSSI}{Received Signal Strength Indicator}
\newacronym{rt}{RT}{Ray Tracer}
\newacronym{rts}{RTS}{Request to Send}
\newacronym{rtt}{RTT}{Round Trip Time}
\newacronym{rw}{RW}{Receive Window}
\newacronym{rx}{RX}{Receiver}
\newacronym{sa}{SA}{standalone}
\newacronym{sack}{SACK}{Selective Acknowledgment}
\newacronym{sap}{SAP}{Service Access Point}
\newacronym{sc}{SC}{Single Carrier}
\newacronym{sch}{SCH}{Secondary Cell Handover}
\newacronym{scm}{SCM}{Spatial Channel Model}
\newacronym{scoot}{SCOOT}{Split Cycle Offset Optimization Technique}
\newacronym{sdma}{SDMA}{Spatial Division Multiple Access}
\newacronym{sdr}{SDR}{Software Defined Radio}
\newacronym{semm}{SEMM}{SPCA-EDCA Mixed Mode}
\newacronym{si}{SI}{Study Item}
\newacronym{sib}{SIB}{Secondary Information Block}
\newacronym{sinr}{SINR}{Signal-to-Interference-plus-Noise Ratio}
\newacronym{sir}{SIR}{Signal-to-Interference Ratio}
\newacronym{sls}{SLS}{Sector-Level Sweep}
\newacronym{sm}{SM}{Saturation Mode}
\newacronym{snr}{SNR}{Signal-to-Noise Ratio}
\newacronym{son}{SON}{Self-Organizing Network}
\newacronym{sp}{SP}{Service Period}
\newacronym{spr}{SPR}{Service Period Request}
\newacronym{srs}{SRS}{Sounding Reference Signal}
\newacronym{ss}{SS}{Synchronization Signal}
\newacronym{sss}{SSS}{Secondary Synchronization Signal}
\newacronym{ssw}{SSW}{Sector Sweep}
\newacronym{sta}{STA}{Station}
\newacronym{stb}{STB}{Set Top Box}
\newacronym{tb}{TB}{Transport Block}
\newacronym[firstplural=Traffic Categories (TCs)]{tc}{TC}{Traffic Category}
\newacronym{tbtt}{TBTT}{Target Beacon Transmission Time}
\newacronym{tcp}{TCP}{Transmission Control Protocol}
\newacronym{tdd}{TDD}{Time Division Duplexing}
\newacronym{tdma}{TDMA}{Time Division Multiple Access}
\newacronym{tfl}{TfL}{Transport for London}
\newacronym{tgad}{TGad}{Task Group ad}
\newacronym{tgay}{TGay}{Task Group ay}
\newacronym{tsconst}{TSCONST}{Traffic Scheduling Constraint}
\newacronym{tsf}{TSF}{Timing Synchronization Function}
\newacronym{tm}{TM}{Transparent Mode}
\newacronym{trp}{TRP}{Transmitter Receiver Pair}
\newacronym{ts}{TS}{Traffic Stream}
\newacronym{tspec}{TSPEC}{Traffic Specification}
\newacronym{tti}{TTI}{Transmission Time Interval}
\newacronym{ttt}{TTT}{Time-to-Trigger}
\newacronym{tx}{TX}{Transmitter}
\newacronym[firstplural=Transmission Opportunities (TXOPs)]{txop}{TXOP}{Transmission Opportunity}
\newacronym{udp}{UDP}{User Datagram Protocol}
\newacronym{ue}{UE}{User Equipment}
\newacronym{ul}{UL}{Uplink}
\newacronym{um}{UM}{Unacknowledged Mode}
\newacronym{uma}{UMa}{Urban Macro}
\newacronym{uml}{UML}{Unified Modeling Language}
\newacronym{up}{UP}{User Priority}
\newacronym{utc}{UTC}{Urban Traffic Control}
\newacronym{vbr}{VBR}{Variable Bit Rate}
\newacronym{vm}{VM}{Virtual Machine}
\newacronym{vr}{VR}{Virtual Reality}
\newacronym{wbf}{WBF}{Wired Bias Function}
\newacronym{wf}{WF}{Wired-first}
\newacronym{wifi}{Wi-Fi}{Wireless Fidelity}
\newacronym{wigig}{WiGig}{Wireless Gigabit}
\newacronym{wlan}{WLAN}{Wireless Local Area Network}
\newacronym{xr}{XR}{eXtended Reality}

%% file: myTikz.tex
\pgfplotsset{compat=newest}
\pgfplotsset{plot coordinates/math parser=false}
\pgfplotsset{every axis/.append style={
                    label style={font=\footnotesize},
                    tick label style={font=\footnotesize},
                    legend style={font=\footnotesize}
                    }}
\usetikzlibrary{plotmarks,shapes,patterns,decorations.pathreplacing,backgrounds,calc,arrows,arrows.meta,spy,matrix,shadows,trees,positioning}
\usepgfplotslibrary{patchplots,groupplots}

\tikzstyle{startstop} = [rectangle, rounded corners, minimum width=2cm, minimum height=0.5cm,text centered, draw=black]
\tikzstyle{io} = [trapezium, trapezium left angle=70, trapezium right angle=110, minimum width=3cm, minimum height=1cm, text centered, draw=black]
\tikzstyle{process} = [rectangle, minimum width=2cm, minimum height=0.5cm, text centered, draw=black, alignb=center]
\tikzstyle{decision} = [ellipse, minimum width=2cm, minimum height=1cm, text centered, draw=black]
\tikzstyle{arrow} = [thick,<->,>=stealth]
\tikzstyle{line} = [thick,>=stealth]
\tikzstyle{darrow} = [thick,<->,>=stealth,dashed]
\tikzstyle{sarrow} = [thick,->,>=stealth]
\tikzstyle{larrow} = [line width=0.1mm,dashdotted,->,>=stealth]

\makeatletter
\def\grd@save@target#1{%
  \def\grd@target{#1}}
\def\grd@save@start#1{%
  \def\grd@start{#1}}
\tikzset{
  grid with coordinates/.style={
    to path={%
      \pgfextra{%
        \edef\grd@@target{(\tikztotarget)}%
        \tikz@scan@one@point\grd@save@target\grd@@target\relax
        \edef\grd@@start{(\tikztostart)}%
        \tikz@scan@one@point\grd@save@start\grd@@start\relax
        \draw[minor help lines] (\tikztostart) grid (\tikztotarget);
        \draw[major help lines] (\tikztostart) grid (\tikztotarget);
        \grd@start
        \pgfmathsetmacro{\grd@xa}{\the\pgf@x/1cm}
        \pgfmathsetmacro{\grd@ya}{\the\pgf@y/1cm}
        \grd@target
        \pgfmathsetmacro{\grd@xb}{\the\pgf@x/1cm}
        \pgfmathsetmacro{\grd@yb}{\the\pgf@y/1cm}
        \pgfmathsetmacro{\grd@xc}{\grd@xa + \pgfkeysvalueof{/tikz/grid with coordinates/major step x}}
        \pgfmathsetmacro{\grd@yc}{\grd@ya + \pgfkeysvalueof{/tikz/grid with coordinates/major step y}}
        \foreach \x in {\grd@xa,\grd@xc,...,\grd@xb}
        \node[anchor=north] at (\x,\grd@ya) {\pgfmathprintnumber{\x}};
        \foreach \y in {\grd@ya,\grd@yc,...,\grd@yb}
        \node[anchor=east] at (\grd@xa,\y) {\pgfmathprintnumber{\y}};
      }
    }
  },
  minor help lines/.style={
    help lines,
    gray,
    line cap =round,
    xstep=\pgfkeysvalueof{/tikz/grid with coordinates/minor step x},
    ystep=\pgfkeysvalueof{/tikz/grid with coordinates/minor step y}
  },
  major help lines/.style={
    help lines,
    line cap =round,
    line width=\pgfkeysvalueof{/tikz/grid with coordinates/major line width},
    xstep=\pgfkeysvalueof{/tikz/grid with coordinates/major step x},
    ystep=\pgfkeysvalueof{/tikz/grid with coordinates/major step y}
  },
  grid with coordinates/.cd,
  minor step x/.initial=.5,
  minor step y/.initial=.2,
  major step x/.initial=1,
  major step y/.initial=1,
  major line width/.initial=1pt,
}
\makeatother

%% file: img/legend_acquisitions.tex
\definecolor{color0}{rgb}{0.42,0.643,0.827}
\definecolor{color1}{rgb}{0.686,0.078,0.086}
\definecolor{color2}{rgb}{.004,.494,.314}
\definecolor{color3}{rgb}{1,0.6,0.2}
\begin{tikzpicture}
\pgfplotsset{every tick label/.append style={font=\scriptsize}}

\begin{axis}[%
width=0,
height=0,
at={(0,0)},
scale only axis,
xmin=0,
xmax=0,
xtick={},
ymin=0,
ymax=0,
ytick={},
axis background/.style={fill=white},
legend style={legend cell align=center, align=center, draw=white!15!black, font=\scriptsize, at={(0, 0)}, anchor=center, /tikz/every even column/.append style={column sep=2em}},
legend columns=5,
]

\addplot [very thick, red]
table {%
0 0
};
\addlegendentry{Expected}
\addplot [only marks, mark=x, very thick, mark size=3, very thick, mark size=3]
table {%
0 0
};
\addlegendentry{Empirical 30 FPS}
\addplot [only marks]
table {%
0 0
};
\addlegendentry{Empirical 60 FPS}

\end{axis}
\end{tikzpicture}%

%% file: img/data_rate.tex
% This file was created by tikzplotlib v0.9.8.
\begin{tikzpicture}

\definecolor{color0}{rgb}{0.42,0.643,0.827}
\definecolor{color1}{rgb}{0.686,0.078,0.086}
\definecolor{color2}{rgb}{.004,.494,.314}
\definecolor{color3}{rgb}{1,0.6,0.2}

\begin{axis}[
width=\fwidth,
height=\fheight,
legend cell align={left},
legend style={
  fill opacity=0.8,
  draw opacity=1,
  text opacity=1,
  at={(0.03,0.97)},
  anchor=north west,
  draw=white!80!black
},
tick align=outside,
tick pos=left,
x grid style={white!69.0196078431373!black},
xlabel={Target Data Rate [Mbps]},
xtick={10,20,30,40,50},
xmin=7.75, xmax=52.25,
xmajorgrids,
xtick style={color=black},
y grid style={white!69.0196078431373!black},
ylabel={Empirical Data Rate [Mbps]},
minor y tick num=1,
ymajorgrids,
yminorgrids,
ymin=-2.51686290816894, ymax=60,
ytick style={color=black}
]
\addplot [very thick, red]
table {%
10 10
20 20
30 30
40 40
50 50
};
% \addlegendentry{Expected}
\addplot [only marks, mark=x, very thick, mark size=3]
table {%
10 10.559504823659
20 20.8074825505727
30 31.1137642274894
40 39.342774241092
50 49.2918489517624
};
% \addlegendentry{Empirical 30 FPS}
\addplot [only marks]
table {%
10 10.9669747779693
20 21.0531788051872
30 31.3873819891688
40 41.5546350937299
50 51.5288060373959
};
% \addlegendentry{Empirical 60 FPS}
\end{axis}

\end{tikzpicture}

%% file: img/non_video_traffic.tex
% This file was created by tikzplotlib v0.9.8.
\begin{tikzpicture}

    \definecolor{color0}{rgb}{0.42,0.643,0.827}
    \definecolor{color1}{rgb}{0.686,0.078,0.086}
    \definecolor{color2}{rgb}{.004,.494,.314}
    \definecolor{color3}{rgb}{1,0.6,0.2}
    
    \begin{axis}[
    width=\fwidth,
    height=\fheight,
    legend cell align={left},
    legend style={
      fill opacity=0.8,
      draw opacity=1,
      text opacity=1,
      at={(0.03,0.5)},
      anchor=west,
      draw=white!80!black
    },
    tick align=outside,
    tick pos=left,
    x grid style={white!69.0196078431373!black},
    xlabel={Target Data Rate [Mbps]},
    xtick={10,20,30,40,50},
    xmin=7.75, xmax=52.25,
    xmajorgrids,
    xtick style={color=black},
    y grid style={white!69.0196078431373!black},
    ylabel={Empirical Data Rate [kbps]},
    ymajorgrids,
    ymin=-2.51686290816894, ymax=150,
    ytick style={color=black}
    ]
    % 30 FPS
    \addplot [only marks, mark=x, very thick, mark size=3, color2, forget plot]
    table {%
    10 2.52629555364159
    20 2.52641930776321
    30 2.52488631016234
    40 2.52480823314016
    50 2.52243641259753
    };
    % \addlegendentry{DL}
    \addplot [only marks, mark=x, very thick, mark size=3, color3, forget plot]
    table {%
    10 111.444856257255
    20 111.860164523769
    30 112.146311510557
    40 112.235348944168
    50 111.580509147643
    };
    % \addlegendentry{UL}

    % 60 FPS
    \addplot [only marks, color2, forget plot]
    table {%
    10 5.4996891255783
    20 4.92545678485712
    30 4.92341733020041
    40 4.92618193719352
    50 4.9271697303368
    };
    % \addlegendentry{Non-video DL 60 FPS}
    \addplot [only marks, color3, forget plot]
    table {%
    10 118.567361028876
    20 118.620497354816
    30 119.188634524455
    40 119.778050773539
    50 119.863838616442
    };
    % \addlegendentry{Total UL 60 FPS}

    \addplot [very thick, color3]
    table {%
    0 0
    };
    \addlegendentry{Uplink}
    \addplot [very thick, color2]
    table {%
    0 0
    };
    \addlegendentry{Downlink}

    \end{axis}
    \end{tikzpicture}
    

%% file: img/frame_size.tex
% This file was created by tikzplotlib v0.9.8.
\begin{tikzpicture}

\definecolor{color0}{rgb}{0.42,0.643,0.827}
\definecolor{color1}{rgb}{0.686,0.078,0.086}

\begin{axis}[
width=\fwidth,
height=\fheight,
legend cell align={left},
legend style={
  fill opacity=0.8,
  draw opacity=1,
  text opacity=1,
  at={(0.5,1.02)},
  anchor=south,
  draw=white!80!black,
  /tikz/every even column/.append style={column sep=1em}
},
legend columns=2,
tick align=outside,
tick pos=left,
x grid style={white!69.0196078431373!black},
xlabel={Empirical Data Rate [Mbps]},
xmajorgrids,
xmin=7.75, xmax=52.25,
xtick style={color=black},
y grid style={white!69.0196078431373!black},
ylabel={Avg. Frame Size [kB]},
ymajorgrids,
ymin=-6.3385955947638, ymax=244.777174156707,
ytick style={color=black}
]
% 30 FPS
\addplot [very thick, red]
table {%
10.5569785281053 43.9874105337722
20.8049561312649 86.6873172136039
31.1112393411792 129.63016392158
39.3402494328588 163.917705970245
49.2893265153498 205.372193813957
};
% \addlegendentry{Ideal}
\addplot [only marks, mark=x, very thick, mark size=3, black]
table {%
10.5569785281053 44.0078516726404
20.8049561312649 86.7205144385027
31.1112393411792 129.746409776923
39.3402494328588 163.911769096491
49.2893265153498 205.365051610961
};
% \addlegendentry{Empirical 30 FPS}

% 60 FPS
\addplot [very thick, red, forget plot]
table {%
10.9614750888437 22.8364064350911
21.0482533484024 43.8505278091716
31.3824585718386 65.3801220246637
41.5497089117927 86.5618935662348
51.5238788676656 107.341414307637
};
% \addlegendentry{Ideal}
\addplot [only marks, black]
table {%
10.9614750888437 22.8679333754627
21.0482533484024 43.870099525359
31.3824585718386 65.4304050833284
41.5497089117927 86.5655545032166
51.5238788676656 107.343929147208
};
% \addlegendentry{Empirical 60 FPS}
\end{axis}

\end{tikzpicture}

%% file: img/ifi.tex
% This file was created by tikzplotlib v0.9.8.
\begin{tikzpicture}

\definecolor{color0}{rgb}{0.42,0.643,0.827}
\definecolor{color1}{rgb}{0.686,0.078,0.086}

\begin{axis}[
width=\fwidth,
height=\fheight,
legend cell align={left},
legend style={fill opacity=0.8, draw opacity=1, text opacity=1, draw=white!80!black,
  at={(0.5,1.02)},
  anchor=south,
  /tikz/every even column/.append style={column sep=1em}
  },
legend columns=2,
tick align=outside,
tick pos=left,
x grid style={white!69.0196078431373!black},
xlabel={Target Data Rate [Mbps]},
xmajorgrids,
xmin=7.75, xmax=52.25,
xtick style={color=black},
y grid style={white!69.0196078431373!black},
ylabel={IFI [ms]},
ymajorgrids,
ymin=0, ymax=40,
ytick style={color=black}
]
% 30 FPS
\addplot [very thick, red]
table {%
10 33.3333333333333
20 33.3333333333333
30 33.3333333333333
40 33.3333333333333
50 33.3333333333333
};
% \addlegendentry{Ideal}
\addplot [only marks, mark=x, very thick, mark size=3, black]
table {%
10 33.3497131250373
20 33.3466217838255
30 33.3633182200842
40 33.3340588731038
50 33.3331991086485
};
% \addlegendentry{Empirical 30 FPS}

% 60 FPS
\addplot [very thick, red, forget plot]
table {%
10 16.6666666666667
20 16.6666666666667
30 16.6666666666667
40 16.6666666666667
50 16.6666666666667
};
% \addlegendentry{Ideal}
\addplot [only marks, black]
table {%
10 16.6900377068914
20 16.6741343662308
30 16.6796418267077
40 16.6673000744602
50 16.6675529180796
};
% \addlegendentry{Empirical 60 FPS}
\end{axis}

\end{tikzpicture}

%% file: img/legend_fit.tex
\definecolor{color0}{rgb}{0.42,0.643,0.827}
\definecolor{color1}{rgb}{0.686,0.078,0.086}
\definecolor{color2}{rgb}{.004,.494,.314}
\definecolor{color3}{rgb}{1,0.6,0.2}
\begin{tikzpicture}
\pgfplotsset{every tick label/.append style={font=\scriptsize}}

\begin{axis}[%
width=0,
height=0,
at={(0,0)},
scale only axis,
xmin=0,
xmax=0,
xtick={},
ymin=0,
ymax=0,
ytick={},
axis background/.style={fill=white},
legend style={legend cell align=center, align=center, draw=white!15!black, font=\scriptsize, at={(0, 0)}, anchor=center, /tikz/every even column/.append style={column sep=2em}},
legend columns=5,
]

\addplot [very thick, dashed]
table {%
0 0
};
\addlegendentry{Fit}
\addplot [only marks, mark=x, very thick, mark size=3, very thick, mark size=3]
table {%
0 0
};
\addlegendentry{Empirical 30 FPS}
\addplot [only marks]
table {%
0 0
};
\addlegendentry{Empirical 60 FPS}
\addplot [very thick, color0]
table {%
0 0
};
\addlegendentry{P-Frames}
\addplot [very thick, color1]
table {%
0 0
};
\addlegendentry{I-Frames}

\end{axis}
\end{tikzpicture}%

%% file: img/ifi_stdev.tex
% This file was created by tikzplotlib v0.9.8.
\begin{tikzpicture}

\definecolor{color0}{rgb}{0.42,0.643,0.827}

\begin{axis}[
width=\fwidth,
height=\fheight,
legend cell align={left},
legend style={fill opacity=0.8, draw opacity=1, text opacity=1, draw=white!80!black},
tick align=outside,
tick pos=left,
x grid style={white!69.0196078431373!black},
xlabel={Target Data Rate [Mbps]},
xmin=7.75, xmax=52.25,
xtick style={color=black},
xtick={10,20,30,40,50},
xmajorgrids,
y grid style={white!69.0196078431373!black},
ylabel={IFI Std. Dev. [ms]},
ymin=0, ymax=4,
ytick style={color=black},
ymajorgrids,
]
% 30 FPS
\addplot [only marks, mark=x, very thick, mark size=3, black, forget plot]
table {%
10 2.6317824017606
20 2.56105719415433
30 3.46654024326962
40 2.24841302947683
50 3.01951451874595
};
\addplot [very thick, black, dashed]
table {%
10 2.78546147748147
20 2.78546147748147
30 2.78546147748147
40 2.78546147748147
50 2.78546147748147
};
% \addlegendentry{mean}

% 60 FPS
\addplot [only marks, forget plot]
table {%
10 1.07886086101991
20 1.56961490057991
30 2.02200570237717
40 1.15447643913214
50 0.997805600699546
};
\addplot [very thick, black, dashed]
table {%
10 1.36455270076174
20 1.36455270076174
30 1.36455270076174
40 1.36455270076174
50 1.36455270076174
};
% \addlegendentry{mean}
\end{axis}

\end{tikzpicture}

%% file: img/gmm_means_S.tex
% This file was created by tikzplotlib v0.9.8.
\begin{tikzpicture}

\definecolor{color0}{rgb}{0.42,0.643,0.827}
\definecolor{color1}{rgb}{0.686,0.078,0.086}
\definecolor{color2}{rgb}{.004,.494,.314}
\definecolor{color3}{rgb}{0.83921568627451,0.152941176470588,0.156862745098039}

\begin{axis}[
width=\fwidth,
height=\fheight,
tick align=outside,
tick pos=left,
x grid style={white!69.0196078431373!black},
xlabel={S [kB]},
xmin=0, xmax=220,
xtick style={color=black},
xmajorgrids,
y grid style={white!69.0196078431373!black},
ylabel={GMM Means [kB]},
ymin=0, ymax=275.762472570755,
ytick style={color=black},
ymajorgrids,
]
% 30 FPS
\addplot [only marks, color0, mark=x, very thick, mark size=3]
table{%
x  y
43.9874105337722 38.3099398432646
86.6873172136039 75.3621382096053
129.63016392158 116.8767014488
163.917705970245 155.371598377985
205.372193813957 185.133147956916
};
\addplot [color1, only marks, mark=x, very thick, mark size=3]
table{%
x  y
43.9874105337722 57.1603452983317
86.6873172136039 111.200691584276
129.63016392158 156.203591402781
163.917705970245 198.091152535039
205.372193813957 231.192662315705
};

% 60 FPS
\addplot [color0, only marks]
table{%
x  y
22.8364064350911 16.3679950604384
43.8505278091716 42.3421354409146
65.3801220246637 58.4879794685096
86.5618935662348 72.9713214252094
107.341414307637 93.0135180167357
};
\addplot [color1, only marks]
table{%
x  y
22.8364064350911 38.3364094872465
43.8505278091716 48.5009129460361
65.3801220246637 75.5736656432016
86.5618935662348 105.512369058656
107.341414307637 113.066754325084
};

% fit
\addplot [very thick, color0, dashed]
table {%
22.8364064350911 20.5703439018064
205.372193813957 184.993057757543
};
\addplot [very thick, color1, dashed]
table {%
22.8364064350911 26.8656600208309
205.372193813957 241.608046012865
};
\end{axis}

\end{tikzpicture}

%% file: img/gmm_stdevs_S.tex
% This file was created by tikzplotlib v0.9.8.
\begin{tikzpicture}

\definecolor{color0}{rgb}{0.42,0.643,0.827}
\definecolor{color1}{rgb}{0.686,0.078,0.086}
\definecolor{color2}{rgb}{.004,.494,.314}
\definecolor{color3}{rgb}{0.83921568627451,0.152941176470588,0.156862745098039}

\begin{axis}[
width=\fwidth,
height=\fheight,
tick align=outside,
tick pos=left,
x grid style={white!69.0196078431373!black},
xlabel={S [kB]},
xmin=20, xmax=250,
xmode=log,
xtick style={color=black},
xtick={1, 2, 5, 10, 20, 50, 100, 200, 500},
xticklabels={1, 2, 5, 10, 20, 50, 100, 200, 500},
extra x ticks={3,4,6,7,8,9,30,40,60,70,80,90,300,400},
extra x tick labels={},
xmajorgrids,
xminorgrids,
y grid style={white!69.0196078431373!black},
ylabel={GMM stdevs [kB]},
ymin=2.5, ymax=50,
ytick={1, 2, 5, 10, 20, 50, 100, 200, 500},
yticklabels={1, 2, 5, 10, 20, 50, 100, 200, 500},
extra y ticks={3,4,6,7,8,9,30,40,60,70,80,90,300,400},
extra y tick labels={},
ymode=log,
ytick style={color=black},
ymajorgrids,
yminorgrids,
]
% 30 FPS
\addplot [only marks, color0, mark=x, very thick, mark size=3]
table{%
x  y
43.9874105337722 8.78062163105488
86.6873172136039 10.7696212187506
129.63016392158 13.2905672687238
163.917705970245 13.665482567833
205.372193813957 20.0809448204423
};
\addplot [only marks, color1, mark=x, very thick, mark size=3]
table{%
x  y
43.9874105337722 9.13570895591042
86.6873172136039 18.243739870626
129.63016392158 19.5763926448485
163.917705970245 36.0308919945953
205.372193813957 33.9536209373628
};

% 60 FPS
\addplot [color0, only marks]
table{%
x  y
22.8364064350911 4.46881019471742
43.8505278091716 7.27437177726914
65.3801220246637 7.71319524919357
86.5618935662348 12.4129295174141
107.341414307637 15.120497810535
};
\addplot [color1, only marks]
table{%
x  y
22.8364064350911 8.51219774691596
43.8505278091716 15.8454837278307
65.3801220246637 14.1064636468249
86.5618935662348 16.6970012109305
107.341414307637 14.3853057618974
};

% fit
\addplot [very thick, color0, dashed]
table {%
22.8364064350911 4.79507494815927
205.372193813957 18.9277190020939
};
\addplot [very thick, color1, dashed]
table {%
22.8364064350911 8.23861803753007
205.372193813957 29.002574564542
};
\end{axis}

\end{tikzpicture}

%% file: img/gmm_probs.tex
% This file was created by tikzplotlib v0.9.8.
\begin{tikzpicture}

\definecolor{color0}{rgb}{0.42,0.643,0.827}
\definecolor{color1}{rgb}{0.686,0.078,0.086}
\definecolor{color2}{rgb}{.004,.494,.314}
\definecolor{color3}{rgb}{0.83921568627451,0.152941176470588,0.156862745098039}

\begin{axis}[
width=\fwidth,
height=\fheight,
tick align=outside,
tick pos=left,
x grid style={white!69.0196078431373!black},
xlabel={Target Rate [Mbps]},
xmin=7.75, xmax=52.25,
xtick style={color=black},
xmajorgrids,
y grid style={white!69.0196078431373!black},
ylabel={GMM weight},
ymin=0, ymax=1,
ytick style={color=black},
ymajorgrids,
]
% 30 FPS
\addplot [only marks, color0, mark=x, very thick, mark size=3]
table{%
x  y
10 0.697730012070065
20 0.683068227945658
30 0.672750417254369
40 0.800087550373193
50 0.5607443123164
};
\addplot [only marks, color1, mark=x, very thick, mark size=3]
table{%
x  y
10 0.302269987929904
20 0.316931772054333
30 0.32724958274562
40 0.199912449626829
50 0.439255687683608
};

% 60 FPS
\addplot [color0, only marks]
table{%
x  y
10 0.704123466138991
20 0.751904646144876
30 0.593670073075407
40 0.582243533424725
50 0.28538162568219
};
\addplot [color1, only marks]
table{%
x  y
10 0.295876533860957
20 0.248095353855044
30 0.406329926924588
40 0.417756466575289
50 0.714618374317717
};

% fit
\addplot [very thick, color0, dashed]
table {%
10 0.640039913732591
50 0.640039913732591
};
\addplot [very thick, color1, dashed]
table {%
10 0.359960086267409
50 0.359960086267409
};
\end{axis}

\end{tikzpicture}

%% file: img/cdfs_ifi_legend.tex
\definecolor{color0}{rgb}{0.42,0.643,0.827}
\definecolor{color1}{rgb}{0.686,0.078,0.086}
\definecolor{color2}{rgb}{.004,.494,.314}
\begin{tikzpicture}
\pgfplotsset{every tick label/.append style={font=\scriptsize}}

\begin{axis}[%
width=0,
height=0,
at={(0,0)},
scale only axis,
xmin=0,
xmax=0,
xtick={},
ymin=0,
ymax=0,
ytick={},
axis background/.style={fill=white},
legend style={legend cell align=center, align=center, draw=white!15!black, font=\scriptsize, at={(0, 0)}, anchor=center, /tikz/every even column/.append style={column sep=2em}},
legend columns=3,
]

\addplot [very thick, color0]
table {%
0 0
};
\addlegendentry{Model}
\addplot [very thick, color1, dashed]
table {%
0 0
};
\addlegendentry{Trace 50 Mbps}
\addplot [very thick, color2, densely dotted]
table {%
0 0
};
\addlegendentry{Trace 10 Mbps}

\end{axis}
\end{tikzpicture}%

%% file: img/cdfs_frame_legend.tex
\definecolor{color0}{rgb}{0.42,0.643,0.827}
\definecolor{color1}{rgb}{0.686,0.078,0.086}
\begin{tikzpicture}
\pgfplotsset{every tick label/.append style={font=\scriptsize}}

\begin{axis}[%
width=0,
height=0,
at={(0,0)},
scale only axis,
xmin=0,
xmax=0,
xtick={},
ymin=0,
ymax=0,
ytick={},
axis background/.style={fill=white},
legend style={legend cell align=center, align=center, draw=white!15!black, font=\scriptsize, at={(0, 0)}, anchor=center, /tikz/every even column/.append style={column sep=2em}},
legend columns=2,
]

\addplot [very thick, color0]
table {%
0 0
};
\addlegendentry{Model}
\addplot [very thick, color1, dashed]
table {%
0 0
};
\addlegendentry{Trace}

\end{axis}
\end{tikzpicture}%

%% file: img/appRate/legend.tex
\definecolor{color0}{rgb}{0.42,0.643,0.827}
\definecolor{color1}{rgb}{0.686,0.078,0.086}
\definecolor{color2}{rgb}{.004,.494,.314}
\begin{tikzpicture}
\pgfplotsset{every tick label/.append style={font=\scriptsize}}

\begin{axis}[%
width=0,
height=0,
at={(0,0)},
scale only axis,
xmin=0,
xmax=0,
xtick={},
ymin=0,
ymax=0,
ytick={},
axis background/.style={fill=white},
legend style={legend cell align=center, align=center, draw=white!15!black, font=\scriptsize, at={(0, 0)}, anchor=center, /tikz/every even column/.append style={column sep=2em}},
legend columns=5,
]

\addplot [very thick, color0]
table {%
0 0
};
\addlegendentry{Model}
\addplot [very thick, color1]
table {%
0 0
};
\addlegendentry{Trace}
% \addplot [very thick, color2]
% table {%
% 0 0
% };
% \addlegendentry{Deterministic}

\addplot [very thick, black]
table {%
0 0
};
\addlegendentry{Burst-level metrics}
\addplot [very thick, black, dashed]
table {%
0 0
};
\addlegendentry{Fragment-level metrics}

\end{axis}
\end{tikzpicture}%

%% file: img/appRate/delay_avg.tex
% This file was created by tikzplotlib v0.9.8.
\begin{tikzpicture}

\definecolor{color0}{rgb}{0.42,0.643,0.827}
\definecolor{color1}{rgb}{0.686,0.078,0.086}
\definecolor{color2}{rgb}{.004,.494,.314}

\begin{axis}[
width=\fwidth,
height=\fheight,
legend cell align={left},
legend style={
  fill opacity=0.8,
  draw opacity=1,
  text opacity=1,
  at={(0.03,0.97)},
  anchor=north west,
  draw=white!80!black
},
tick align=outside,
tick pos=left,
x grid style={white!69.0196078431373!black},
xlabel={Target Data Rate [Mbps]},
xmajorgrids,
xmin=-0.45, xmax=4.45,
xtick style={color=black},
xtick={0,1,2,3,4},
xticklabels={10,20,30,40,50},
y grid style={white!69.0196078431373!black},
ylabel={Avg. Delay [ms]},
ymajorgrids,
ymin=0, ymax=1.5,
ytick style={color=black}
]

% burst avg
\path [draw=color0, very thick]
(axis cs:0,0.267868100286265)
--(axis cs:0,0.269358206524573);

\path [draw=color0, very thick]
(axis cs:1,0.50111455611131)
--(axis cs:1,0.504046845721907);

\path [draw=color0, very thick]
(axis cs:2,0.802597727755747)
--(axis cs:2,0.808382304213747);

\path [draw=color0, very thick]
(axis cs:3,1.12728846971666)
--(axis cs:3,1.13201140335468);

\path [draw=color0, very thick]
(axis cs:4,1.39677373770749)
--(axis cs:4,1.40337615429774);

\path [draw=color1, very thick]
(axis cs:0,0.292213186903596)
--(axis cs:0,0.292635618428928);

\path [draw=color1, very thick]
(axis cs:1,0.520467292526879)
--(axis cs:1,0.520876602036051);

\path [draw=color1, very thick]
(axis cs:2,0.841225308363808)
--(axis cs:2,0.842117477385826);

\path [draw=color1, very thick]
(axis cs:3,1.15517021889086)
--(axis cs:3,1.15617955238376);

\path [draw=color1, very thick]
(axis cs:4,1.43556106954412)
--(axis cs:4,1.43660210523817);

% \path [draw=color2, very thick]
% (axis cs:0,0.268979622115873)
% --(axis cs:0,0.269219867630302);

% \path [draw=color2, very thick]
% (axis cs:1,0.492954766428472)
% --(axis cs:1,0.493177088816441);

% \path [draw=color2, very thick]
% (axis cs:2,0.913288210675261)
% --(axis cs:2,0.914448569453508);

% \path [draw=color2, very thick]
% (axis cs:3,1.13728732877089)
% --(axis cs:3,1.13844527124519);

% \path [draw=color2, very thick]
% (axis cs:4,1.36137011978458)
% --(axis cs:4,1.36253663927615);

\addplot [very thick, color0]
table {%
0 0.268613153405419
1 0.502580700916609
2 0.805490015984747
3 1.12964993653567
4 1.40007494600262
};
\addplot [very thick, color1]
table {%
0 0.292424402666262
1 0.520671947281465
2 0.841671392874817
3 1.15567488563731
4 1.43608158739115
};
% \addplot [very thick, color2]
% table {%
% 0 0.269099744873088
% 1 0.493065927622457
% 2 0.913868390064384
% 3 1.13786630000804
% 4 1.36195337953036
% };

% fragment avg
\path [draw=color0, very thick]
(axis cs:0,0.177118716681836)
--(axis cs:0,0.178116295274195);

\path [draw=color0, very thick]
(axis cs:1,0.293418185713438)
--(axis cs:1,0.294940743625315);

\path [draw=color0, very thick]
(axis cs:2,0.425688361981194)
--(axis cs:2,0.428190555152717);

\path [draw=color0, very thick]
(axis cs:3,0.57489674152375)
--(axis cs:3,0.577822568132686);

\path [draw=color0, very thick]
(axis cs:4,0.722532943053994)
--(axis cs:4,0.726078706614014);

\path [draw=color1, very thick]
(axis cs:0,0.20511450501908)
--(axis cs:0,0.205873397110923);

\path [draw=color1, very thick]
(axis cs:1,0.305944839921344)
--(axis cs:1,0.306456357914983);

\path [draw=color1, very thick]
(axis cs:2,0.444132954390773)
--(axis cs:2,0.444573052539509);

\path [draw=color1, very thick]
(axis cs:3,0.603931531700145)
--(axis cs:3,0.604521247322847);

\path [draw=color1, very thick]
(axis cs:4,0.743080007447076)
--(axis cs:4,0.743725121060274);

% \path [draw=color2, very thick]
% (axis cs:0,0.168632539512919)
% --(axis cs:0,0.169282197785794);

% \path [draw=color2, very thick]
% (axis cs:1,0.279077378450661)
% --(axis cs:1,0.279472710559082);

% \path [draw=color2, very thick]
% (axis cs:2,0.39441531099341)
% --(axis cs:2,0.394734989948723);

% \path [draw=color2, very thick]
% (axis cs:3,0.553035368313896)
% --(axis cs:3,0.55345521731096);

% \path [draw=color2, very thick]
% (axis cs:4,0.692852498415305)
% --(axis cs:4,0.693423393326786);

\addplot [very thick, color0, dashed]
table {%
0 0.177617505978015
1 0.294179464669377
2 0.426939458566955
3 0.576359654828218
4 0.724305824834004
};
\addplot [very thick, color1, dashed]
table {%
0 0.205493951065002
1 0.306200598918164
2 0.444353003465141
3 0.604226389511496
4 0.743402564253675
};
% \addplot [very thick, color2, dashed]
% table {%
% 0 0.168957368649357
% 1 0.279275044504871
% 2 0.394575150471067
% 3 0.553245292812428
% 4 0.693137945871046
% };
\end{axis}

\end{tikzpicture}

%% file: img/appRate/delay_95.tex
% This file was created by tikzplotlib v0.9.8.
\begin{tikzpicture}

\definecolor{color0}{rgb}{0.42,0.643,0.827}
\definecolor{color1}{rgb}{0.686,0.078,0.086}
\definecolor{color2}{rgb}{.004,.494,.314}

\begin{axis}[
width=\fwidth,
height=\fheight,
legend cell align={left},
legend style={
  fill opacity=0.8,
  draw opacity=1,
  text opacity=1,
  at={(0.03,0.97)},
  anchor=north west,
  draw=white!80!black
},
tick align=outside,
tick pos=left,
x grid style={white!69.0196078431373!black},
xlabel={Target Data Rate [Mbps]},
xmajorgrids,
xmin=-0.45, xmax=4.45,
xtick style={color=black},
xtick={0,1,2,3,4},
xticklabels={10,20,30,40,50},
y grid style={white!69.0196078431373!black},
ylabel={Delay 95\textsuperscript{th} pctl [ms]},
ymajorgrids,
ymin=0, ymax=2.1,
ytick style={color=black}
]
% burst 95
\path [draw=color0, very thick]
(axis cs:0,0.398819231587702)
--(axis cs:0,0.403567180412298);

\path [draw=color0, very thick]
(axis cs:1,0.731283545148735)
--(axis cs:1,0.779671338851265);

\path [draw=color0, very thick]
(axis cs:2,1.21126691614803)
--(axis cs:2,1.22044876185197);

\path [draw=color0, very thick]
(axis cs:3,1.50081497570569)
--(axis cs:3,1.51208633629431);

\path [draw=color0, very thick]
(axis cs:4,1.99434288652108)
--(axis cs:4,2.00622198947892);

\path [draw=color1, very thick]
(axis cs:0,0.582184280536672)
--(axis cs:0,0.585198207463328);

\path [draw=color1, very thick]
(axis cs:1,0.668780638444758)
--(axis cs:1,0.670576837555242);

\path [draw=color1, very thick]
(axis cs:2,1.22638577981533)
--(axis cs:2,1.23157928418467);

\path [draw=color1, very thick]
(axis cs:3,1.57805504923055)
--(axis cs:3,1.58419836276945);

\path [draw=color1, very thick]
(axis cs:4,1.96336693347746)
--(axis cs:4,1.96936882252254);

% \path [draw=color2, very thick]
% (axis cs:0,0.27255213383957)
% --(axis cs:0,0.27258905816043);

% \path [draw=color2, very thick]
% (axis cs:1,0.496550081791027)
% --(axis cs:1,0.496578202208973);

% \path [draw=color2, very thick]
% (axis cs:2,0.976423073234405)
% --(axis cs:2,0.977181882765595);

% \path [draw=color2, very thick]
% (axis cs:3,1.20038021515105)
% --(axis cs:3,1.20117876084895);

% \path [draw=color2, very thick]
% (axis cs:4,1.42453284538992)
% --(axis cs:4,1.42525061061008);

\addplot [very thick, color0]
table {%
0 0.401193206
1 0.755477442
2 1.215857839
3 1.506450656
4 2.000282438
};
\addplot [very thick, color1]
table {%
0 0.583691244
1 0.669678738
2 1.228982532
3 1.581126706
4 1.966367878
};
% \addplot [very thick, color2]
% table {%
% 0 0.272570596
% 1 0.496564142
% 2 0.976802478
% 3 1.200779488
% 4 1.424891728
% };

% fragment 95
\path [draw=color0, very thick]
(axis cs:0,0.328867116290947)
--(axis cs:0,0.332157201709053);

\path [draw=color0, very thick]
(axis cs:1,0.567244500021315)
--(axis cs:1,0.571805665978685);

\path [draw=color0, very thick]
(axis cs:2,1.00413010062615)
--(axis cs:2,1.01022747337385);

\path [draw=color0, very thick]
(axis cs:3,1.23955286173371)
--(axis cs:3,1.24690037226629);

\path [draw=color0, very thick]
(axis cs:4,1.47038451065374)
--(axis cs:4,1.47906818534626);

\path [draw=color1, very thick]
(axis cs:0,0.481869066049147)
--(axis cs:0,0.482798053950853);

\path [draw=color1, very thick]
(axis cs:1,0.563589307236385)
--(axis cs:1,0.564118282763615);

\path [draw=color1, very thick]
(axis cs:2,1.03223577395298)
--(axis cs:2,1.03390836404702);

\path [draw=color1, very thick]
(axis cs:3,1.32041521954221)
--(axis cs:3,1.32252744245779);

\path [draw=color1, very thick]
(axis cs:4,1.49054403318501)
--(axis cs:4,1.49242412481499);

% \path [draw=color2, very thick]
% (axis cs:0,0.266625139599082)
% --(axis cs:0,0.266674644400918);

% \path [draw=color2, very thick]
% (axis cs:1,0.484624055128247)
% --(axis cs:1,0.484736080871753);

% \path [draw=color2, very thick]
% (axis cs:2,0.693107584709299)
% --(axis cs:2,0.693287623290701);

% \path [draw=color2, very thick]
% (axis cs:3,1.1074434962166)
% --(axis cs:3,1.1088281497834);

% \path [draw=color2, very thick]
% (axis cs:4,1.3172006204689)
% --(axis cs:4,1.3184762115311);

\addplot [very thick, color0, dashed]
table {%
0 0.330512159
1 0.569525083
2 1.007178787
3 1.243226617
4 1.474726348
};
\addplot [very thick, color1, dashed]
table {%
0 0.48233356
1 0.563853795
2 1.033072069
3 1.321471331
4 1.491484079
};
% \addplot [very thick, color2, dashed]
% table {%
% 0 0.266649892
% 1 0.484680068
% 2 0.693197604
% 3 1.108135823
% 4 1.317838416
% };
\end{axis}

\end{tikzpicture}

%% file: img/nStas/legend.tex
\definecolor{color0}{rgb}{0.42,0.643,0.827}
\definecolor{color1}{rgb}{0.686,0.078,0.086}
\definecolor{color2}{rgb}{.004,.494,.314}
\begin{tikzpicture}
\pgfplotsset{every tick label/.append style={font=\scriptsize}}

\begin{axis}[%
width=0,
height=0,
at={(0,0)},
scale only axis,
xmin=0,
xmax=0,
xtick={},
ymin=0,
ymax=0,
ytick={},
axis background/.style={fill=white},
legend style={legend cell align=center, align=center, draw=white!15!black, font=\scriptsize, at={(0, 0)}, anchor=center, /tikz/every even column/.append style={column sep=2em}},
legend columns=5,
]

\addplot [very thick, color0]
table {%
0 0
};
\addlegendentry{Model}
\addplot [very thick, color1]
table {%
0 0
};
\addlegendentry{Trace}
% \addplot [very thick, color2]
% table {%
% 0 0
% };
% \addlegendentry{Deterministic}

\addplot [very thick, black]
table {%
0 0
};
\addlegendentry{60 FPS}
\addplot [very thick, black, dash dot]
table {%
0 0
};
\addlegendentry{30 FPS}

\end{axis}
\end{tikzpicture}%

%% file: img/nStas/delay_avg.tex
% This file was created by tikzplotlib v0.9.8.
\begin{tikzpicture}

\definecolor{color0}{rgb}{0.42,0.643,0.827}
\definecolor{color1}{rgb}{0.686,0.078,0.086}
\definecolor{color2}{rgb}{.004,.494,.314}

\begin{axis}[
width=\fwidth,
height=\fheight,
legend cell align={left},
legend style={
  fill opacity=0.8,
  draw opacity=1,
  text opacity=1,
  at={(0.03,0.97)},
  anchor=north west,
  draw=white!80!black
},
tick align=outside,
tick pos=left,
x grid style={white!69.0196078431373!black},
xlabel={\# STAs},
xmajorgrids,
xmin=0.7, xmax=8.3,
minor x tick num=1,
xminorgrids,
xtick style={color=black},
y grid style={white!69.0196078431373!black},
ylabel={Avg. Delay [ms]},
ymajorgrids,
ymin=0, ymax=8,
ytick style={color=black}
]

\addplot [very thick, color0]
table {%
1 1.40007494600262
2 1.51703077625978
3 1.65943739592508
4 1.83012348424993
5 2.03524095259255
6 2.3362279646669
7 2.75485061812291
8 3.54910442623485
};
\addplot [very thick, color1]
table {%
1 1.43608158739115
2 1.55385109686543
3 1.71928320631018
4 1.90869336099593
5 2.15537930429155
6 2.39854982624159
7 2.83741938819023
8 3.71582975547548
};

\addplot [very thick, color0, dash dot]
table {%
1 2.86102005713003
2 3.11216182933542
3 3.39120210695318
4 3.76412792376
5 4.28594241579881
6 4.8495520639036
7 5.69604772063547
8 7.02534419283536
};
\addplot [very thick, color1, dash dot]
table {%
1 2.57481714780953
2 2.97555665307565
3 3.17107150929342
4 3.44778333247122
5 3.91838902862588
6 4.53441478408478
7 5.25958727489428
8 6.32269105308381
};
% \addplot [very thick, color2, dashed]
% table {%
% 1 2.8828369422563
% 2 3.11069320491148
% 3 3.37253516211802
% 4 3.69217703817609
% 5 3.99170830869788
% 6 5.03361568541571
% 7 6.93796273614934
% 8 7.8845642570352
% };

\end{axis}

\end{tikzpicture}

%% file: img/nStas/delay_95.tex
% This file was created by tikzplotlib v0.9.8.
\begin{tikzpicture}

\definecolor{color0}{rgb}{0.42,0.643,0.827}
\definecolor{color1}{rgb}{0.686,0.078,0.086}
\definecolor{color2}{rgb}{.004,.494,.314}

\begin{axis}[
width=\fwidth,
height=\fheight,
legend cell align={left},
legend style={
  fill opacity=0.8,
  draw opacity=1,
  text opacity=1,
  at={(0.03,0.97)},
  anchor=north west,
  draw=white!80!black
},
tick align=outside,
tick pos=left,
x grid style={white!69.0196078431373!black},
xlabel={\# STAs},
xmajorgrids,
xmin=0.7, xmax=8.3,
minor x tick num=1,
xminorgrids,
xtick style={color=black},
y grid style={white!69.0196078431373!black},
ylabel={Delay 95\textsuperscript{th} pctl [ms]},
ymajorgrids,
ymin=0, ymax=20,
ytick style={color=black}
]

\addplot [very thick, color0]
table {%
1 2.000282438
2 2.448372777
3 3.03781608
4 3.6542018
5 4.357087662
6 5.392289846
7 6.810298697
8 9.459208752
};
\addplot [very thick, color1]
table {%
1 1.966367878
2 2.350029944
3 2.921834859
4 3.584965387
5 4.391243465
6 5.115202147
7 6.365301137
8 9.149928425
};

\addplot [very thick, color0, dash dot]
table {%
1 3.833882713
2 5.017507455
3 6.075960094
4 7.279544435
5 9.051639566
6 10.762486728
7 13.330989326
8 17.231641335
};
\addplot [very thick, color1, dash dot]
table {%
1 3.792623182
2 4.871653609
3 5.758933265
4 6.697360827
5 8.145159237
6 10.039427615
7 12.169709104
8 15.463405293
};
% \addplot [very thick, color2, dashed]
% table {%
% 1 3.009963356
% 2 3.474281007
% 3 4.227161883
% 4 5.293774095
% 5 6.311246846
% 6 9.108486399
% 7 13.162638819
% 8 15.122616812
% };

\end{axis}

\end{tikzpicture}